\pdfoutput=1

\documentclass[acmlarge, nonacm]{acmart}

\AtBeginDocument{%
  \providecommand\BibTeX{{%
    \normalfont B\kern-0.5em{\scshape i\kern-0.25em b}\kern-0.8em\TeX}}}

\acmJournal{POMACS}
\acmVolume{37}
\acmNumber{4}
\acmArticle{111}
\acmMonth{8}

\usepackage{subcaption}
\usepackage{wrapfig}

\usepackage{enumitem}

\setcopyright{none}

\begin{document}

\settopmatter{printacmref=false}
\settopmatter{printfolios=false}
\renewcommand\footnotetextcopyrightpermission[1]{} 
\pagestyle{plain} 

\title{COVID-19 Contact Tracing and Privacy: A Longitudinal Study of Public Opinion}




\def\reportversion{2.0}

\author{Lucy Simko}
\affiliation{%
  \institution{Paul G. Allen School of Computer Science \& Engineering, University of Washington}}

\author{Jack Lucas Chang}
\affiliation{%
  \institution{Information School, University of Washington}}
  
\author{Maggie Jiang}
\affiliation{%
  \institution{Paul G. Allen School of Computer Science \& Engineering, University of Washington}}
  
\author{Ryan Calo}
\affiliation{%
\institution{School of Law, University of Washington}}

\author{Franziska Roesner}
\affiliation{%
 \institution{Paul G. Allen School of Computer Science \& Engineering, University of Washington}}
  
\author{Tadayoshi Kohno}
\affiliation{%
 \institution{Paul G. Allen School of Computer Science \& Engineering, University of Washington}}

\begin{abstract}
There is growing use of technology-enabled contact tracing, the process of identifying potentially infected COVID-19 patients by notifying all recent contacts of an infected person. Governments, technology companies, and research groups alike have been working towards releasing smartphone apps, using IoT devices, and distributing wearable technology to automatically track ``close contacts'' and identify prior contacts in the event an individual tests positive. However, there has been significant public discussion about the tensions between effective technology-based contact tracing and the privacy of individuals. To inform this discussion, we present the results of seven months of online surveys focused on contact tracing and privacy, each with 100 participants. Our first surveys were on April 1 and 3, before the first peak of the virus in the US, and we continued to conduct the surveys weekly for 10 weeks (through June), and then fortnightly through November, adding topical questions to reflect current discussions about contact tracing and COVID-19. Our results present the diversity of public opinion and can inform policy makers, technologists, researchers, and public health experts on whether and how to leverage technology to reduce the spread of COVID-19, while considering potential privacy concerns. We are continuing to conduct longitudinal measurements and will update this report over time; citations to this version of the report should reference Report Version 2.0, December 4, 2020.
\end{abstract}

\pagestyle{plain}


\maketitle

\section{Introduction}

Technology companies, university research groups, and governments have been diligently working to deploy COVID-19 contact tracing apps, for which adoption has been slow~\cite{singaporeDownloadRate, delagarzaWhyArent}. Prior work has determined that contact tracing apps are most effective when used by the majority of a population~\cite{forbesContactTracing, ferretti2020quantifying, not60pct}; however, some have raised security and privacy concerns (e.g.,~\cite{acluPrivacyConcerns, effProximityAppsChallenges}) as well as broader concerns about efficacy (e.g.,~\cite{efficacyCTapps}). 

Our research seeks to provide to the scientific, technology, and policy communities an informed understanding of the public's values, concerns, and opinions about the use of automated contact tracing technologies. We argue neither for nor against automated contact tracing in this work, but instead we offer a summary of public opinion on potential contact tracing scenarios since many regions have already implemented automated contact tracing programs or are moving towards them. We ask the following research questions:
\begin{itemize}[leftmargin=*]
  \item \textbf{App functionality.} What do potential users want a contact tracing app to do or not do? What data sources do people feel most and least comfortable with being used for contact tracing? Our survey asks about potential app features and multiple data sources, including: location data (e.g., from cell tower data, credit card history, or wearable electronics), proximity data, data from an existing app, and data from a new app by a known entity or company (Sections~\ref{section:likelihood} and~\ref{section:alternate-sources}).
  \item \textbf{Developer and stakeholder identity.} What kinds of institutions do potential users trust to conduct or implement automated contact tracing? We ask about trust in a number of potential developers, including government agencies and well known tech companies (Section~\ref{section:who-develops}). We also solicit individuals' opinions regarding contact tracing data being shared with or used by different entities for the purposes of contact tracing. We consider data sharing with and usage by multiple entities, including: their government, cellular provider, cellphone manufacturer, and various well-known technology companies (Sections~\ref{section:who-develops} and ~\ref{section:alternate-sources}).
  \item \textbf{Changes over time.} How, if at all, has public opinion about the preceding topics changed over time? We discuss longitudinal changes in Sections~\ref{section:likelihood},~\ref{section:who-develops}, and~\ref{section:alternate-sources}. We also ask whether there are any correlations with demographic factors or world events, e.g., the global or regional infection rate (Section~\ref{section:demographic-trends}). 
\end{itemize}

We capture public opinion using an international paid survey platform (Prolific). Our first survey (April 1, 2020) was repeated weekly through June and fortnightly thereafter, with the latest data collected on November 6, 2020. Our first survey collected data from 200 participants, and the rest collected data from 100 participants each. Our first surveys were before the initial viral peak infection rate in North America, before contact tracing apps were available in many of the regions that now have them, and early in the public discourse about contact tracing; our later surveys track how public opinion evolves over time. 

This paper addresses a broad audience---researchers, app developers, public health officials, policy makers, etc. Our results can inform (1) ongoing technical efforts to design contact tracing apps in a privacy-preserving manner, (2) how the makers of such a contact tracing app or program communicate the privacy properties of their contact tracing program to their potential users, and (3) legal, ethical and policy discussions about the appropriate use and design of such technologies. At a high level, we find: 
\begin{itemize}[leftmargin=*]
    \item \textbf{Privacy preferences are stable over time at a population level.} We find that public opinion about privacy and contact tracing is roughly stable over time, suggesting that our---and others'---results can successfully inform future efforts. We find there is a shrinking population that has yet to use contact tracing apps, and that privacy concerns limit potential users' willingness to download the apps. Therefore, the population that does not yet use a contact tracing app may appear to become more privacy conscious as the less privacy conscious leave it and download an app (Section~\ref{section:likelihood}).
    \item \textbf{An abundance of privacy concerns about data sharing, usage, and developer identity leads to a personal decision about the tradeoffs between privacy and health, and leaves no perfect solution.} 
    We observe that potential contact tracing app users care deeply about the identity of the developer, and have strong opinions about with whom data should or should not be shared. However, we note that participants disagree about trusted entities. Many participants raised concerns about sharing with their government and data being used for advertising or government surveillance, now or in the future (Section~\ref{section:who-develops}).

    \item \textbf{Informed consent and transparency about data sharing and usage may mitigate some privacy concerns.} Participants expressed a strong desire for meaningful consent and control over their data. If developers and policy makers (1) better inform the public about the current and future use of their data, and (2) give individuals control over how their data are used, they may be more willing to enroll in automated contact tracing. For example, we find support for judicial oversight of government data usage in some circumstances, potentially making users more confident that their data would not be misused (Section~\ref{section:who-develops}).
    
    \item \textbf{Mental models of technical and legal concepts are often incomplete or inaccurate, but play a significant role in potential users' willingness to begin contact tracing.} Participants repeatedly reasoned about the \textit{accuracy} of certain technical methods of contact tracing (e.g., GPS vs Bluetooth), the \textit{competence} of the app developer to implement contact tracing at a technical level, and the \textit{capability} of their government to protect (or exploit) their data. Through this reasoning, we identified multiple inaccurate or incomplete mental models, e.g., some participants thought a proximity tracking app would be less secure than a location tracking app due to constant communication with others' phones via Bluetooth. Other participants overestimated the prevalence of judicial corruption, causing them to discount the protection potentially provided by judicial oversight of government data usage. These mental models invite stakeholders to improve user education so that users can make well-informed decisions.

\end{itemize}

\section{The evolution of contact tracing during the COVID-19 pandemic}

\begin{figure}[t]
    \includegraphics[scale=.7]{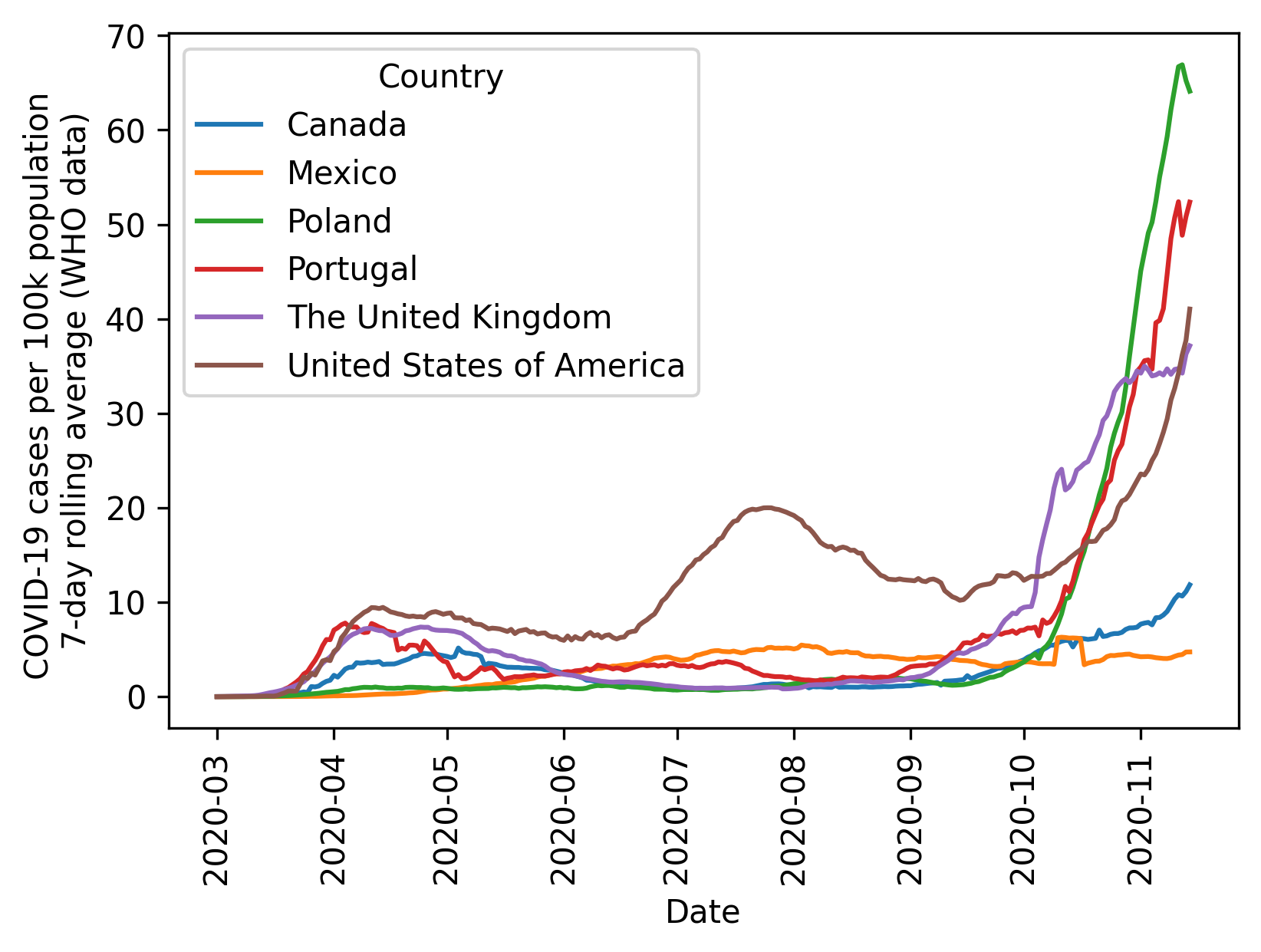}
    \caption{COVID-19 new infections per 100k as reported to the World Health Organization (WHO)~\cite{whoinfectiondata} for the six countries from which we had at least 100 participants (together, these countries comprise 74.4\% of our participant pool).
}
    \label{fig:world-covid}
\end{figure}

Because both the pace of development of contact tracing apps and the pandemic itsef are evolving rapidly, we expect content in this section to be quickly outdated. However, it is important to capture the state of the world on April 1, when we first deployed our survey, and see how both the infection rates and contact tracing efforts have progressed through the time of this writing in mid-November, in order to contextualize our results. 

\subsection{COVID-19 infection rates and quarantine restrictions}

\hspace{\parindent}\textbf{Infection rates.} On April 1, the course of the COVID-19 pandemic had not yet reached its first peaks outside of Asia. Figure~\ref{fig:world-covid} shows the number of infections per 100k people in the six countries from which we had at least 100 participants total: Canada, Mexico, Poland, Portugal, the UK, and the US. After an initial spring peak in many countries in our dataset, the rate of infection declined. The US saw a second peak of infection in August, though rates of infection in many others countries remained low~\cite{europeDelineRate}. As of this writing, in mid-November, many countries are experiencing skyrocketing infection rates, as seen in Figure~\ref{fig:world-covid}.

\textbf{Regional lockdowns.} On April 1, as we began our survey, many European countries, (e.g., the UK, Germany, Italy, and Spain), were under varying forms of lockdown, with some combination of schools, restaurants, bars, and non-essential shops closed, public gatherings banned, and citizens urged or mandated to stay inside except for essential outings~\cite{germanyContractTracingApril1, march31, march30Spain, ukLockdown}. Many in the US were under similar restrictions, though some states issued no stay-at-home orders at all during those early months~\cite{April1NYT, usStatesApril1, usStatesApril1_2, nytSeeWhichStates}.

Due to the lower infection rates over the summer, restrictions largely eased in Europe but have been re-implemented in many countries as of mid-November in the form of nightly curfews, closures of non-essential businesses, travel restrictions, and mask-wearing and social distancing mandates  ~\cite{europeLockdownNov, europeTravelRestrictionsNov, europeMasksNov}. As of mid November, restrictions in many US states are not as strict, with limitations but not bans on indoor activities (such as dining and shopping) and masks mandated in some but not all states~\cite{nytSeeWhichStatesNov14}.

\subsection{Contact tracing technical efforts and app adoption}

Here, we briefly overview existing contact tracing app efforts and their adoption as well as the conversation around how to contact trace in a privacy-preserving way. The purpose of this section is to contextualize our findings and recommendations, not to give a comprehensive look at technology-enabled contact tracing efforts.  

\hspace{\parindent}\textbf{Why automated contact tracing?} Traditionally, contact tracing is done by a team of public health experts and focuses on tracking down those who might have been infected by someone who tested positive for a disease, in combination with widespread testing. A state, region, or other entity might implement \textit{automated} contact tracing (e.g., to augment or complement human-based efforts, for which there has been a shortage~\cite{lackOfCT1, lackofCT2}) for multiple reasons, although though not all experts agree that automated contact tracing is needed or will be effective. For example, automated contact tracing might be used to keep infection rates low while allowing people to leave their homes, or used to enforce quarantine for people identified as COVID-19-positive.

\hspace{\parindent}\textbf{Existing automated contact tracing programs.} As of April 1, some governments had already deployed automated contact tracing programs using a variety of devices and data sources~\cite{wikipediaCovidApps}. For example, contact tracing apps existed in Bahrain, China, Colombia, the Czech Republic, Ghana, India, Israel, the Republic of North Macedonia, Norway, Singapore, and some US States~\cite{chinaContactTracing, traceTogether, bahrainApp, colombiaApp, czechApp, ghanaApp, indiaApp, israelContactTracing, israelApp, macedoniaApp, norwayApp, utahContactTracing, northDakotaContactTracing}. Some apps were mandatory (e.g., in China), but most were optional (e.g., Singapore) and struggled with low adoption~\cite{singaporeDownloadRate}. Hong Kong deployed electronic wristbands to those infected with COVID-19 to ensure they did not leave their homes~\cite{hongKongWristbands}. In South Korea, the government sent text messages with details about new COVID-19 cases and made available a central database with anonymized information; however, some entries were specific enough to be traced back to a single person and initiated damaging rumors~\cite{southKoreaContactTracing, southKoreaCoronaMap}. Additionally, Taiwan and Israel both began using cell tower data~\cite{israelContactTracing, taiwanGeofencing}.

On April 10, Google and Apple announced ``\textit{a joint effort to enable the use of Bluetooth technology to help governments and health agencies reduce the spread of the virus, with user privacy and security central to the design}''~\cite{appleGoogleAnnouncement}. In May, they released the first version of their Exposure Notification API~\cite{googleENUpdateJuly, appleENReleaseDate, macrumorsEN}. The API uses proximity tracking through Bluetooth, is opt-in, and can be used only by public health authorities;  Apple and Google report that the data will not be monetized~\cite{googleENUpdateJuly, macrumorsEN}. In September, Apple and Google launched ``Exposure Notification Express,'' allowing users to participate in contact tracing without downloading an app~\cite{appleENReleaseDate}.

Since the release of Apple and Google's Exposure Notification API, many public health authorities have released apps using the API. According to the \textit{MIT Technology Review} COVID Tracing Tracker~\cite{MITTRtrackingtracker}, as of mid-November, 46 non-US countries are using automated contact tracing applications, 13 with Apple and Google's API and 4 with  DP-3T~\cite{dp3t}. Additionally, at least 12 US States are using the Apple and Google API as of mid-November~\cite{ENwikipedia,qzUSStatesENApps}. Apps have been released steadily over time around the world, yet adoption remains low in most regions: of the apps for which the adoption rate is in the \textit{MIT Technology Review'}s database, as of mid-November, Iceland, Ireland, and Singapore have the highest voluntary adoption rates, at just below 40\% of the population of each country~\cite{MITTRtrackingtracker}. Though automated contact tracing is more effective with more users, it can be effective at low rates of adoption as well~\cite{not60pct}.

\textbf{Design decisions affecting security and privacy.} Design properties at multiple levels affect user security and privacy, some of which are transparent to users (e.g., being potentially identified as infectious in some designs) and some of which are more opaque  (e.g., broadcast vs narrowcast; centralization vs decentralization). For a more complete and in-depth discussion of these properties, see~\cite{redmilesUserConcerns, centralizedOrDecentralized}. 

Some designs explicitly focus on privacy-respecting contact tracing. Each group makes design decisions based on its own threat models and on-the-ground situations to produce different technical protocols, including Apple and Google, the Massachusetts Institute of Technology (MIT), the University of Washington (UW), Inria, PEPP-PT, and DP-3T, and PanCast~\cite{appleGoogleAnnouncement, mitPact, uwPact, ROBERT, dp3t, barthe2020pancast}. One high-level distinction that has become extremely popular since our initial surveys in early April is \textit{proximity tracking}, where a user's phone tracks other nearby phones, rather than a more traditional implementation of location-based contact tracing.

Additionally, other, non-smartphone methods are being used to trace contacts, such as credit card purchase history, facial recognition on surveillance camera footage, and wearable devices~\cite{russiaFacialRecognition, singerAsCoronavirusNYT, wearableCT}.

\section{Related Work} \label{section:related-work}

Other groups have also investigated public opinion on location tracking during COVID-19, described below. 

\textbf{Themes: privacy concerns abound, but a majority indicate a willingness to download contact tracing apps.} Many groups have assessed a population's willingness to download a contact tracing app, finding rates between 27 and 84\% at different points in time, with different privacy and data sharing and usage situations and different populations, including Australia~\cite{australiaPublicOpinion, garrett2020acceptability, socialLicensing2020}, China~\cite{kostka2020times}, a number of countries in Western Europe~\cite{hargittaiStudy, altmann2020acceptability, o2020national, jansen2020predictors, wiertz2020predicted, guillon2020attitudes, kostka2020times, williams2020public, abeler2020support}, and the US~\cite{wapoMostAmericans, hargittaiWillAmericans, hargittaiStudy, altmann2020acceptability, redmilesHowGood, li2020decentralized, zhang2020americans, pewMostAmericans, kostka2020times, abeler2020support}. These works identified concerns thematically similar to ours, such as privacy concerns about sharing with the government, and correlations between willingness to download and COVID-19 concern levels or demographic information, e.g., age.  

\textbf{Cross cultural studies.} Some groups have studied participants from multiple countries, including ~\cite{altmann2020acceptability, abeler2020support}. For example, Altmann et al.\ found that people in the US and Germany were less likely than people in France, Italy, and the UK to install a contact tracing app due to security and privacy concerns~\cite{altmann2020acceptability}. Kostka and Habich-Sobiegalla compared public acceptance of contact tracing apps in China, Germany, and the US; in line with Altmann et al., they find that participants in Germany and the US were much less accepting of an app than those in China~\cite{kostka2020times}.  
\textbf{Longitudinal studies.} Garrett et al.\ are studying public opinion over time in several countries (including Australia, Germany, and the UK) by periodically surveying participants from those countries in ``waves''~\cite{socialLicensing2020}. In Australia, Garrett et al. have found widespread acceptance for contact tracing apps but lower download rates than were predicted by attitudes about contact tracing apps~\cite{garrett2020acceptability}.

\textbf{Situating our work.} In the context of existing work, our work adds a \textit{regular and periodic} survey of public opinion, capturing trends and stability over time. Additionally, the free response answers present in our data provide rich insight into the values and concerns underlying individuals' willingness to download, allowing them to express themselves in their own words in addition to via prescribed quantitative options.

\section{Methodology}

To collect rich data and measure public opinion, we designed an approximately 20-minute online survey with both multiple choice and free response questions. Our survey was implemented in Qualtrics. We deployed the survey through Prolific, an online survey platform based in the United Kingdom. 

Our institution's IRB determined that our study was exempt from further human subjects review, and we adhered to best practices for ethical human subjects survey research, e.g., we paid at or slightly above minimum wage, all questions were optional except the initial screening questions about age and smartphone usage, and we did not collect unnecessary personal information. 

\subsection{Survey Protocol}

Because we expected most participants to live in countries where contact tracing apps were not in ubiquitous use, at least for our initial survey, we designed the survey to elicit attitudes about contact tracing in specific \textit{hypothetical} situations. The survey did include branches for those who had already downloaded an app for tracking or mitigating COVID-19, or who had the opportunity to but chose not to, but in this paper, \textit{we focus on those who did not have a contact tracing app} at the time of inquiry. To avoid biasing participants towards presenting themselves as more privacy-conscious than they are, the survey did not mention ``privacy'' until its final two questions (demographics) and asked instead about participants' ``comfort'' with various situations, or their ``likelihood'' of downloading an app in a certain situation. Each section (except for demographics) concluded with one or more free-response questions, inviting participants to explain their opinions. 

When designing this survey in late March---and adding to it in response to the evolving world---we paid close attention to the ways that technology and terminology might change, opting to describe terms that may fall in or out of style (like ``contact tracing'' or ``exposure notification'') and prioritized longitudinal consistency by not editing questions after they had appeared once (other than to correct the rare typo). We expand on this experience of future-proofing a longitudinal survey during a rapidly evolving event in Section~\ref{section:discussion}.

The survey had the following main sections (excluding questions for participants who were already using a contact tracing app). See Appendix~\ref{survey-protocol} for the full protocol. 

\textbf{Demographics.} We asked participants three types of demographic questions that focused variables we hypothesized might correlate with their attitude towards COVID-19 and contact tracing programs: (1) standard demographic questions, like age, gender, geographic location; (2) general political views, news sources, and privacy and technology interest and knowledge; (3) COVID-19-specific questions, like their general level of concern about the pandemic, whether they live with someone who is in a high-risk group, whether they had had COVID-19 or had ever been tested, and their beliefs about social distancing and mask wearing. We asked many of the demographic questions at the end of the survey to help mitigate stereotype threat.

\textbf{Cell tower location data.} We asked participants how comfortable they were with their cell phone manufacturer or cellular carrier using their location data for the purposes of studying or mitigating the spread of COVID-19. We presented participants with three variations of a situation: their location data being shared with their government; their location data being shared with their government if they tested positive; and their location data being shared publicly if they tested positive.

\textbf{Existing apps using GPS location data.} We asked participants to imagine that ``the makers of an existing app on your phone started using your GPS location data to study or mitigate the spread of COVID-19.'' We chose 3 popular apps from each of 5 categories that we expected would use location data (navigation, social media, messaging, transportation, fitness), for a total of 15 apps. Participants rated their comfort level with each of the 15 apps using their location data for mitigating the spread of COVID-19 on a 5-point Likert scale, with an additional option for ``I don't use this app.'' We then asked two free-response questions about the app that they regularly use that they would \textit{most} trust and the app that they would \textit{least} trust to study or mitigate COVID-19.

\textbf{New app: perfect privacy.} We asked participants to imagine a new app that would track their location for the purposes of mitigating the spread of COVID-19 but that would protect their data perfectly. On 5-point Likert scales, we asked how likely they would be to install the app and how it would change their current behavior.

\textbf{New app: app makers know location at all times but do not share it.} Changing the previous scenario slightly, we asked participants to imagine a new app that would know their location at all times for the purposes of mitigating the spread of COVID-19, but this time the app makers would know their location at all times but would not share this information. We again asked participants how likely they would be to download and use such an app. This time, we asked participants to rate their comfort with each company that made the same popular 15 apps we showed them previously. We expanded this list to include other companies in week 3 of the survey. We also asked about their comfort level with five generic entities making such an app: a university research group, an activist group, an industry startup, your government, and the United Nations.

\textbf{New app: app makers know location at all times and share data with your government if you are diagnosed with COVID-19.} Again changing the previous scenarios, we asked participants about a situation in which the new app's makers share their location history with the government if they test positive for COVID-19. We asked, again, how likely they would be to download such an app as well as their download likelihood in two variant situations: if the data were shared regardless of whether they tested positive, and if the government's use of the data were supervised by a judge.

\textbf{Non-smartphone location data sources.} In response to an evolving conversation about alternate data sources, we asked participants next about their comfort level with having location history derived from surveillance camera footage and credit card purchase history (added in week 3). Beginning in week 16, we also asked participants about their comfort with public area sensors or electronic bracelets.

\textbf{New app: proximity tracking.} Due to the growing discussions about and technical work on proximity tracking protocols and apps after April 1, in week 3 we added a group of questions about proximity tracking. We asked about proximity tracking by phone manufacturers, phone operating systems, a new app, and apps from several well known companies or generic entities.

\textbf{Government use of location or proximity data.} In this section, we stepped back from scenarios about specific data sources to ask participants questions about a scenario in which their government acquires their location or proximity data for studying and mitigating COVID-19. We asked about their confidence in their government's deletion of the data post-pandemic, use of data only for COVID-19 tracking, and their general level of concern about their ``personal safety or the safety of those in their community.''

\textbf{Desired features in a new COVID-19 mitigation app.} We then asked participants about a wide variety of features that a potential new COVID-19 mitigation app might have when notifying people of potential infections or enforcing isolation; features were drawn from existing contract tracing apps or programs. For example, one feature we asked about would ``notify you if you came close to someone who later tested positive for COVID-19,'' while another would ``automatically notify the authorities if people were not isolating as mandated.''

\textbf{Location sharing with their government pre-pandemic.} Finally, we asked participants to rate their level of comfort with their location data being shared with their government in October 2019, i.e., before COVID-19. Since participants may not accurately recall their own previous beliefs or may have been primed towards privacy-sensitivity by the rest of the survey. Therefore, any results from this data must be treated with caution.

\subsection{Recruitment}

We recruited participants through Prolific, an online survey platform, with no demographic restrictions, since Prolific already requires that all participants be 18 or older. The first questions of our survey screened participants as required by our IRB. We asked: (1) are you at least 18 years old? and (2) do you use a smartphone regularly? If participants answered `Yes' to both, they proceeded to the rest of the survey.

We ran the survey on Prolific on April 1, 3, 8, 10, and every Friday thereafter until June 5, then every \textit{other} Friday, around the same time (3pm PST). We excluded anyone who had taken any previous version of the survey. 

\subsection{Analysis}

In this report, we present analyses of our qualitative and quantitative data. We conducted exploratory and descriptive statistical analysis of our quantitative data rather than testing specific hypotheses, described below.

\textbf{Longitudinal analysis.} To explore longitudinal trends, we present data with time on the x axis and the percent of participants on the y axis. We draw slopes that are statistically significant with p <= .05, a standard threshold for significance, but we observe that a statistically significant slope does not necessarily mean that the slope has practical significance. We calculate the statistical mean ($\mu$) for each question. 

\textbf{Demographic analysis.} For the questions that displayed longitudinal stability (the majority of the questions), we examined demographic trends by collapsing all weeks of data into one pool. We analyzed each question by: country or region, age bracket, gender, and phone manufacturer, including only demographic groups for which there were at least 100 participants.

\textbf{Qualitative analysis.} To understand participants' values and concerns more deeply, we conducted qualitative analysis of the optional free response questions accompanying many of the survey sections. To analyze these questions, two researchers iteratively created independent qualitative codebooks for each question, first open coding and then creating axial and hierarchical codes for each question. We opted to use separate codebooks for every question except questions that were variants in order to allow the themes from one question to arise independently from the themes in another. When reporting qualitative data, we report the number of participants for each theme, idea, or concept, and attribute quotes to participants using an identifier with both the week and a participant number, e.g., W3P40 for participant 40 from week 3.

\subsection{Limitations}

We are presenting these results without peer review because we believe our results have the potential to inform the current rapidly-evolving policy and technical work on contact tracing apps.  However, because the paper has yet to undergo peer review, and despite our good faith attempts never to mislead or misrepresent our findings, we hope the reader will be especially cautious in interpreting our results.

Online surveys have inherent limitations. Participants may experience survey fatigue and click through long matrix questions, giving inaccurate answers in order to finish the survey more quickly. From our qualitative analysis, responses to free response questions seem to be on topic and high quality, indicating a low rate of survey fatigue. Survey fatigue, or lack thereof, may also be affected by the fact that participants were paid and therefore incentivized to finish. Additionally, though we intentionally recruited from an international audience, our survey was in English, meaning that those who do not read English are not represented, and some with weaker English skills may have chosen not to complete the free response questions; this could lead to potential biases towards English speakers in the qualitative responses.

Another limitation of a survey such as ours, in which participants are asked about different situations without being able to directly compare them, is that opinions about earlier questions may change given later questions (order bias)~\cite{redmiles2017summary}. Thus, because question order matters, our quantitative results in Figures such as Figure~\ref{fig:how-likely} and~\ref{fig:cell_tower_data} must be interpreted with caution, though we believe there is still value in presenting the questions comparatively. 

Prior work on Mechanical Turk participants in the United States, a different survey platform than ours, found, with varied results, that online survey participants may not be representative of the general population~\cite{ross2010crowdworkers}. Other studies have examined whether online survey participants' security and privacy knowledge and behavior accurately represent the general public, with varying results~\cite{kang2014privacy, redmiles2019well}.

Finally, although we have an international sample, we do not recruit large numbers of participants from any individual country each week, so our data cannot be used to examine country-level trends \textit{over time}.

\section{Results}

We now report results from our analysis of both quantitative and qualitative data from weeks 1 (April 1, 2020) through 32 (November 6, 2020) of our survey. For weeks 1-10 (through June 5), we surveyed participants once a week; for subsequent weeks, we surveyed every two weeks (hence, there is no data for weeks 11, 13, 15, etc.). We excluded participants who had participated in any prior version of the survey.

Our survey had two branches, one addressing those who had a contact tracing app and one addressing those who did not. In this paper, we focus on the latter cohort since they may share concerns and values we must understand in order to make it possible for them to (1) have safe access to automated contact tracing, and (2) be able to make an informed decision about participating.

In Section~\ref{section:demographics}, we describe our population demographically, finding that while our participants hail from dozens of countries, minority viewpoints may be absent. In Section~\ref{section:likelihood}, we consider estimates of a population's willingness to download a contact tracing app. We also consider how privacy concerns affect willingness to download. We find that while adoption of contact tracing applications \textit{is} increasing, a significant minority of the population does not intend to use them, and that privacy concerns are indeed a central concern, even amongst those who might download an app. We also consider functionality users might want from contact tracing apps, finding support for bare-bones tracing features but not for more privacy invasive ones, such as quarantine enforcement.

In Section ~\ref{section:who-develops}, we examine values and concerns potential app users might share about tech companies, governments, or other entities that develop contact tracing apps. We observe that users have substantial concerns about their data being shared or used without their consent and for purposes that might harm them or others. We also find no one-size-fits-all app developer profile: comfort with an app developer (e.g., Google or the US government) is a complex decision that differs for every user; therefore, policy makers, tech companies, researchers, governments, and public health experts must work together towards protecting users and helping users understand the protections in place so that they can make informed decision. Finally, in Section~\ref{section:alternate-sources}, we more broadly explore user values and concerns through a discussion of alternative data sources for contact tracing, including cell tower location data, credit card history, public sensors (including surveillance cameras), and wearable electronics. Expanding upon themes from previous sections, we observe that anonymity and technical \textit{accuracy} are of great concern to users, whose mental models may be incomplete or inaccurate.

Some notes on terminology: We use the term `contact tracing' to include `location tracking' and `proximity tracking.' When reporting qualitative results, we use the format W1P100 to mean participant 100 from week 1. Further, our notation Q[$N$], where $N$ is a number, refers to unique identifiers in the Qualtrics survey platform that we used. Question numbers do not appear strictly in order; we analyze questions in groups but encourage readers to refer to the full survey protocol in Appendix~\ref{survey-protocol} if necessary. Finally, in longitudinal plots, we draw lines when statistically significant (p$\leq$.05) and show the average ($\mu$) in the legend for all questions.

\subsection{Participant Demographics: European, male, white, and young} \label{section:demographics}

Over the course of 20 surveys and 32 weeks, we reached \textbf{2337 participants}, \textbf{mostly from Europe}. Countries with at least 10\% of participants in our dataset are the United Kingdom and Northern Ireland (22.4\%),  Portugal (15.9\%), and Poland (14.6\%). 9.9\% of our participants were from United States. European countries (including the UK, Portugal, Poland, and 21 others) comprised 73\% of total survey participants.

As is common in online surveys~\cite{ross2010crowdworkers}, participants were \textbf{overwhelmingly young}. Over 70\% were under the age of 30; 54.9\% between ages 18 and 24 (we screened out anyone younger than 18); 18.9\% between 25 and 29; and 11.2\% between 30 and 34, with a long tail to a highest age bracket of 70-74.

40.7\% of participants who disclosed their gender were female; \textbf{58.1\% were male}. Approximately 1\% of participants disclosed that they were transgender, genderfluid, genderqueer, non-binary, or agender. We manually bucketed participants' gender identities as reported in a free response text field; we believe we have stayed true to participants' gender identities when bucketing, though these identities may change over time and our respondents may have included more trans or gender non-conforming participants than disclosed as such. To avoid stereotype threat, we asked most demographic questions at the end of the survey, asking only high-level questions about demographics (e.g., location) and COVID-19 at the start.

In week 12 (June 19), we began collecting data on race and ethnicity. Because race and ethnicity are complex and have different meanings throughout the world, we provided participants with a number of races or ethnicities commonly asked about in surveys~\cite{fda2016guidance, pewRace} and also offered a free response question if they wished to self describe in addition to or instead of options we provided, as recommended by the EU~\cite{euRace}. Of these responses, \textbf{79.6\% identified as white}, 13.5\% as Hispanic or Latinx, 6.9\% as Asian, 3.2\% as Black or African-American, and less than 1\% as American Indian, Alaskan Native, Pacific Islander, or Native Hawaiian. Percentages add up to more than 100 because some participants selected multiple identities. Those who chose to self describe indicated both intersectional identities and European ethnicities, such as Slavic, Irish, and Scandinavian.

Though we have survey participants from a variety of countries, they are overwhelmingly young, white, and European. Thus, our survey is dominated by racial and ethnic majorities in the countries that we surveyed from, and privacy concerns of those who we were unable to reach---\textbf{older adults, racial and ethnic minorities---are not well represented in our dataset.}

\subsubsection{Participants are concerned about COVID-19 and believe in social distancing and mask wearing}
\begin{figure}[t]
    \includegraphics[scale=.55]{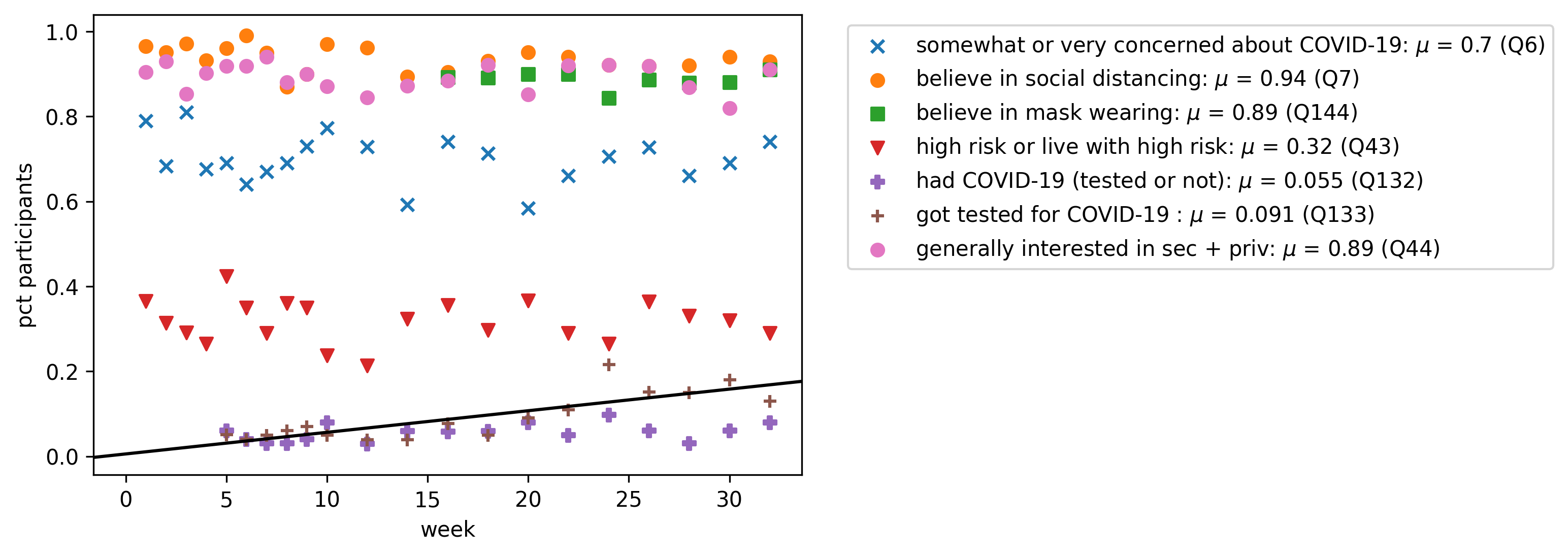}
    \caption{This plot describes our participants' attitudes toward COVID-19 and preventative measures. Respondents have a generally high degree of belief in preventative measures, like mask wearing a social distancing. The line at the bottom shows a statistically significant increase in the percent of participants who had been tested for COVID-19.}
    \label{fig:covid-demographics}
\end{figure}

Figure~\ref{fig:covid-demographics} summarizes COVID-19-related demographic information about participants, revealing no statistically significant longitudinal trend other than a slight increase in those who were tested for COVID-19 (Q133, p<.01). However, our data reveal that our participants are generally concerned about COVID-19 ($\mu$ = 70\%) and believe in social distancing ($\mu$ = 94\%) and mask wearing ($\mu$ = 89\%) as preventative measures for COVID-19. A sizeable minority are in a high risk category or living with someone who is ($\mu$ = 32\%).

A report on public attitudes in the US towards mask wearing and other COVID-19 prevention measures found that mask wearing increased substantially between June and August in many regions of the US~\cite{moreAmericansMaskWearing}. We do not see such an increase in support for mask wearing in our data but do find support for such measures similar to higher numbers from~\cite{moreAmericansMaskWearing}; the lack of trend in our data may be explained by our smaller and Euro-centric population. A May 2020 CDC report found that those surveyed in Los Angeles and New York City overwhelmingly supported mask wearing and social distancing measures~\cite{czeisler2020public}, with numbers similar to our results. However, as revealed in~\cite{moreAmericansMaskWearing}, many US regions showed limited support for mask wearing and other measures: these views are not represented in our data, so our findings must be interpreted carefully to consider those who we did not reach.

\subsection{Expectations about app functionality, data sharing, and technical implementation suggest privacy and accuracy concerns will influence users' willingness to download contact tracing apps}

\label{section:likelihood}

\begin{figure}[t]
\begin{subfigure}[t]{.5\textwidth}
  \centering
  \includegraphics[width=.9\linewidth]{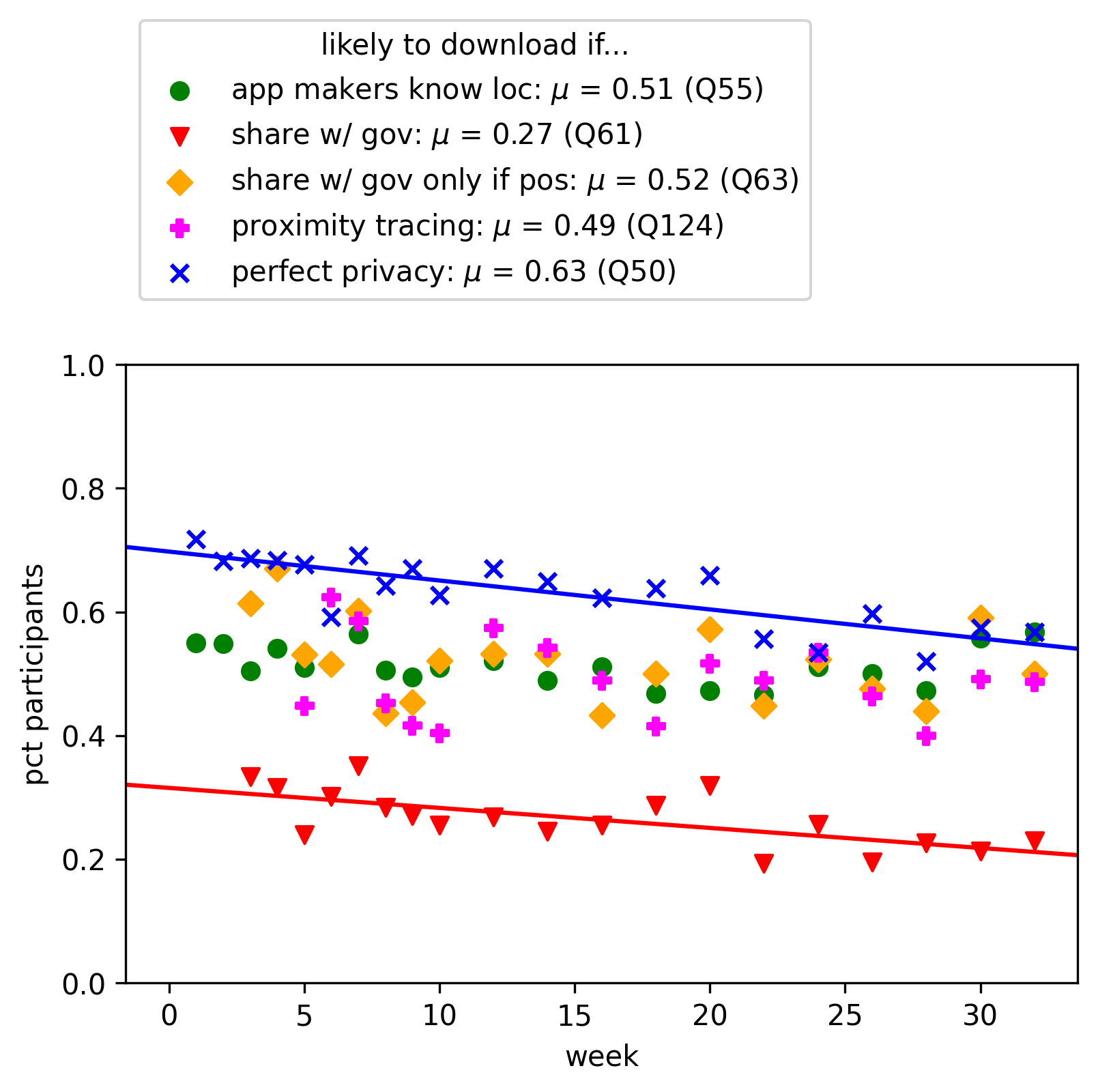}  
  \caption{This graph addresses the question: what kind of app are those who do not already have a contact tracing app willing to download? One might expect this sector of the population---a shrinking portion---to become proportionally more privacy-conscious over time since those who are less privacy-conscious may download an app voluntarily.}
  \label{fig:likely-sub-no-app}
\end{subfigure}~\hspace{15pt}
\begin{subfigure}[t]{.5\textwidth}
  \centering
  \includegraphics[width=.9\linewidth]{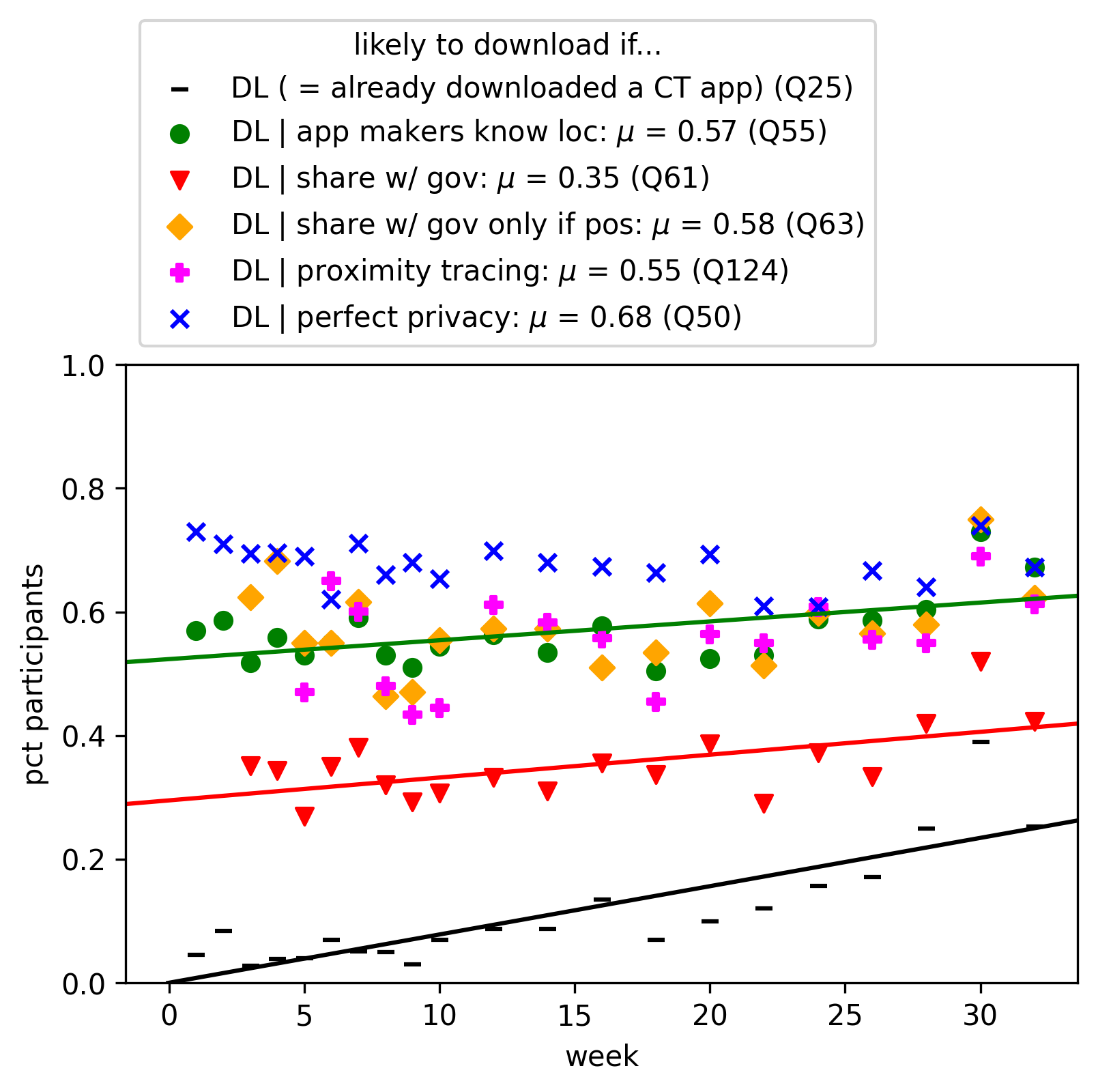}   
  \caption{This graph addresses the question: What percent of the population as a whole might have a contact tracing app in the future? (i.e., each colored point is the sum of those who already have a contact tracing app and those who might be willing to download one)}
  \label{fig:likely-sub-all-ps}
\end{subfigure}
\caption{The percent of respondents who indicated that they would be somewhat or extremely likely to download an app that tracked their location or proximity to others ``for the sake of tracking or mitigating COVID-19. The left plot (a) shows \textit{only the participants who do not have a contact tracing application.} The right plot (b) shows \textit{\textbf{all} participants: each colored point is the \textbf{sum} of those who already have a contact tracing app and those who said would be somewhat or extremely likely to download one.}}
\label{fig:how-likely}
\end{figure}

We now ask what proportion of our population might be willing to download a contact tracing application, and what concerns or values they might have about what the app does. Estimating the number of people willing to download under \textit{some} circumstances is critical to finding the best-case success of voluntarily downloaded contact tracing applications. Understanding potential users' concerns and values will help app makers and public health experts in ongoing efforts to tailor policy, technology, and public awareness campaigns towards reaching global critical mass usage of automated contact tracing. We find that:
\begin{itemize}[leftmargin=*]
    \item The upper bound of people willing to download a contact tracing app remains roughly constant over time, but potential \textbf{users may be becoming more willing to accept an app that does not have perfect privacy} (i.e., an app that shares data under some circumstances) (Sections~\ref{section:not-everyone-will-dl} and~\ref{section:not-everyone-dl}).
    \item When comparing contact tracing apps that use location tracking (GPS) and proximity tracking (Bluetooth), participants considered both \textbf{expected technical accuracy} and \textbf{privacy concerns} (Section~\ref{section:loc-vs-prox}).
    \item Participants value a contact tracing app that \textbf{notifies them (or others) if they have been exposed to COVID-19, but not an app that would enforce isolation or quarantine} (e.g.,~\cite{ctEnforcement}), citing concerns about algorithmic inaccuracy and equity. In qualitative results, participants organically suggested  \textit{informational} features, such as regional guidelines or medical advice, perhaps revealing a lack of reliable information about COVID-19 or an unmet desire for a unified or official source (Sections~\ref{section:features}).
\end{itemize}

\subsubsection{Not everyone intends to download a contact tracing app; privacy concerns reduce likelihood to download, but some concerns may be decreasing over time} \label{section:not-everyone-will-dl}

Figure~\ref{fig:likely-sub-no-app} shows that even if a contract tracing app were to ``protect your data perfectly,'' a significant minority of those who have not yet downloaded an app do not intend to voluntarily do so. Approximately 63\% said they would be somewhat or extremely likely to download a contact tracing app with perfect privacy, while many fewer would download an app that shared their location with their government ($\mu$ = 27\%). Participants showed no significant preference between an app for which the app makers have access to location (Q55), an app for which the developers share data for those who test positive with the government (Q63), or proximity tracing (Q124), all around 50\% of those who have not yet downloaded an app.

Over time, we observe a convergence of those who would download an app with perfect privacy (denoted by the blue `x's in Figure~\ref{fig:how-likely}) and those who would download an app in any of the other circumstances we presented, other than all data being shared with their government (i.e., the other data points except for the red triangles). This convergence may indicate that as contact tracing apps are released, those who have not yet downloaded them are increasingly accepting of ceding some privacy, though they remain concerned about sharing data with their government. This introduces an opportunity for app developers and governments to increase participation in automated contact tracing, while it also imposes a responsibility on policy makers, app developers, and governments to do so responsibly and with respect for users' data.

\subsubsection{Concerns about sharing data with the government limits willingness to download; users may be more likely to cede some privacy if they test positive for COVID-19}\label{section:data-sharing-gov-if-pos}

Participants were more comfortable with an app that would share location data only from users that tested positive ($\mu$ = 52\%) than one that would share location data from all users ($\mu$ = 27\%). This difference in comfort reveals that participants have strong concerns about sharing location data with their government and that those concerns may limit their willingness to download a contact tracing app. This difference in opinion also raises questions about participants' mental models of the mechanics of contact tracing---who do they believe is conducting contact tracing?---as well as their views of government data usage (explored further in Section~\ref{section:who-develops}). 

More broadly, these results suggest that those who test positive for COVID-19 may be more likely to cede some privacy. W1P194 wrote: ``\textit{If I were to be tested positive for the virus, I would definitely sacrifice some of my privacy to the government if it means protecting others. However I’m conflicted on the thought of sharing this data with the government if I am healthy.}'' W6P58 said: ``\textit{I don't want the government to track my location. However, if i tested positive for Covid-19 I understand why it would be necessary so I would reluctantly accept it in that case.}''

\subsubsection{Not everyone will voluntarily download a contact tracing app}\label{section:not-everyone-dl}

In Figure~\ref{fig:likely-sub-all-ps}, we add to Figure~\ref{fig:likely-sub-no-app}---the participants who have already downloaded a contact tracing app---in order to ask a subtly different question: what percentage of smartphone users might have a contact tracing app in the future? We observe, first, that the percent of our participants who have downloaded a contact tracing application is steadily, and significantly, increasing over time (p < .01), from less than 5\% in week 1 to almost 25\% in week 32\footnote{Some participants may have misunderstood the question asking whether they had a contact tracing application and answered `Yes' when they did not have an app. In a cursory estimate, we find that about 10-15\% of participants wrote mainstream apps like ``Google Maps'' instead of a contact tracing app; however, here, we count \textit{all} `yes' answers because it is possible that the participants who answered incorrectly \textit{believe} that they have a contact tracing app, given that the incorrect apps are still likely to ask for location or Bluetooth permissions.}). By adding the data from Figure~\ref{fig:likely-sub-no-app} on top of the participants who already have a contact tracing app, we find that approximately 68\% of our participants have either already downloaded a contact tracing application or would be willing to download one under certain circumstances (including ``an app that protects your data perfectly'').

Estimates vary on how much of a population needs to participate in contact tracing in order for it to halt the pandemic, but recent work suggests a rate of around 60\%~\cite{ferretti2020quantifying, forbesContactTracing} to 70\%~\cite{hellewell2020feasibility}, though ~\cite{ferretti2020quantifying} shows that automated contact tracing at any rate will slow the pandemic. 

Our data show that even in the best possible situation, with ``an app that protects your data perfectly'' (Q50), many participants have reservations about using an app to study or mitigate COVID-19. Results from other surveys about the same topic have shown a willingness to download ranging from 27 to 84\%, depending on the population and exact situations presented. The wide range of willingnesses in related work suggests that further work is necessary to examine the differences between the populations studied and the exact situations presented.

\subsubsection{Location tracking presents privacy and security concerns that proximity does not, reducing appeal to users; participants also reasoned about equity and technical accuracy} \label{section:loc-vs-prox} 

Participants did not exhibit a strong preference for proximity tracking over two other forms of location tracking in our quantitative data (Figure~\ref{fig:how-likely}), potentially due to bias inherent with question ordering since location tracking situations were presented first. However, qualitative data reveals underlying concerns about location tracking compared to proximity tracking, as well as concerns about privacy and efficacy of proximity tracking itself. We also find that inaccurate mental models of technology and contact tracing drive individuals' concerns, values, and willingness to download.

\textbf{Participants have technical security concerns with proximity tracking.} Proximity tracing evoked security concerns, with participants specifically concerned about anonymity and the security risks of their phone communicating directly with others': ``\textit{I don't want the phones to be sharing information between them because it could be easy for a hacker to violate multiple phones privacy}'' (W6P74). W16P59 imagined a scenario in which tracing close contacts instead of location might put political dissidents at risk: ``\textit{Again, there are MANY people for whom this would simply not be safe if our current government had that information. Identify, say, one person at a protest. Get the info of EVERY phone that came within 6 feet of them on that day, or in that timeframe, and suddenly a LOT of people are at risk that had not been identified, and most often had done nothing wrong.}'' Though these scenarios raise questions about the accuracy of participants' mental models of proximity tracing, they reveal that participants' concerns about the technical safety of a contact tracing method will drive their willingness to enroll in automated contact tracing initiatives. Additionally, recent work has shown that many contact tracing apps have suboptimal privacy and security properties~\cite{sharma2020use, wen2020study, sun2020vetting}, so even though participant' mental models were inaccurate, their fears were not unfounded.

\textbf{Qualitative data shows fewer privacy concerns about proximity tracking than about location tracking.} Despite their concerns about privacy, 98 participants wrote a response that indicated they preferred proximity tracking, while only 13 preferred location. 18 indicated that it depended on who ran the service. Those who preferred proximity tracking considered it less invasive than location tracking. W9P100 brought up concerns about contact tracing data being shared with both companies and their government, and reasoned that proximity data was more private: ``\textit{Proximity will probably be easier to swallow than location. There's something fundamentally unsettling about companies / my government having a record of everywhere I've gone, for how long I stayed there, etc. Knowing who I've passed on the street or purchased a burrito from, but not precisely where I passed them or exactly when I bought the burrito would be much less uncomfortable.}'' 

\textbf{Participants reasoned about accuracy and effectiveness of both location tracking and proximity tracking.} This reasoning revealed inaccurate mental models, e.g., ``\textit{I think that proximity tracking is more effective...as it would... be able to alert others and possibly stop them further spreading it, whereas location tracking would only really give an insight into where the virus is spreading}'' (W6P23). Taking the opposite stance, W9P46 wrote: ``\textit{This method [proximity] appears to be better in a user data protection sense. But it does not provide the same benefits to the government that location data would. Location data enables lockdown and focus on specific areas with recent outbreaks.}'' Other participants were concerned about proximity tracing not capturing surface transmission: ``\textit{There is also some evidence about catching from surfaces that others have touched, so location tracking may also be relevant}'' (W14P67). Regardless of the accuracy of participants' mental models---both about the mechanics of automated contact tracing and about virus transmission---their concerns about efficacy reveal that they value a technology that they believe can \textit{accurately} conduct automated contact tracing, and that their views of what technology is most likely to produce accurate contact tracing results will play into their decision to download.

\subsubsection{Participants desire bare-bones contact tracing and informational features, not an app that might enforce quarantine or reveal personal information}\label{section:features}

\begin{figure}
    \includegraphics[scale=.6]{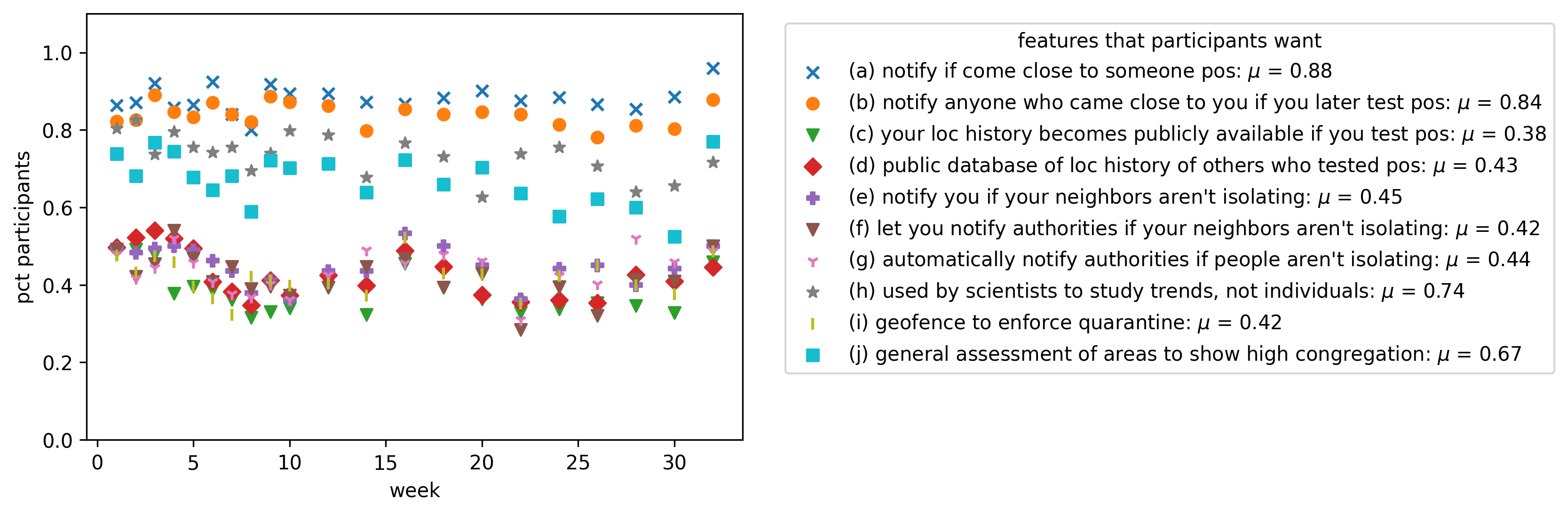}
    \caption{Features desired by participants in a COVID-19 tracking app (Q72).}
    \label{fig:features}
\end{figure}

The features or functionality of an app may also influence how many people are willing to download it. Participants responded positively to four of the ten potential app features\footnote{All features were drawn from existing or proposed designs as of April 1, 2020.}, shown in Figure~\ref{fig:features}: they desire contact tracing notifications (a, b) ($\mu$ = 88\%, 84\%) and general reports on trends (h, j) ($\mu$ = 74\%, 67\%). All other potential features received less than 50\% support.

In qualitative responses (Q74), 106 participants indicated support for informational features, with 44 desiring general information such as news and safety guidelines, and 62 expressing interest in location-specific resources such as information about nearby hospitals. W6P97 suggested that ``\textit{it would be useful to have a hotline or chat where you could be evaluated and diagnosed,}'' and W1P54 thought it would be useful for a contact tracing app to ``\textit{provide national announcements and guidelines so that people get them in a clear, uniform fashion.}'' 31 participants also expanded upon (j), expressing support for an app that shows COVID-19 hotspots, and 39 supported the options to notify (or be notified) if in contact with a positive case (a, b). Participants' desire for an \textit{informative} app raises the question of whether access to reliable information about COVID-19 is an issue.

Concerns about security, privacy, equity, and access also arose. 41 participants mentioned anonymity, specifically bringing up stalking, harassment, and other forms of app abuse. Participants also reiterated their desire for health information to be shared with scientists, health professionals, and family, but not with the public. Though 16 participants wanted enforcement of rules to keep themselves or others safe, others were strongly opposed to this idea, citing concerns about fairness: ``\textit{I would prefer it to only inform and not gather any data or contact any law enforcers because everyone has their own circumstances and there might be people who cannot be on quarantine because of being not wealthy enough}'' (W6P53). W14P75 noted that developers must be mindful of resource consumption and backwards compatibility lest they risk excluding people since ``\textit{not all of us have the privilege of having the latest models}.''


\subsection{App developer identity matters: (Mis)trust in government and some companies} \label{section:who-develops}

We next more deeply investigate the privacy concerns revealed in Section~\ref{section:likelihood} by exploring participants' trust in both government agencies and well-known tech companies. We find that:

\begin{itemize}[leftmargin=*]
    \item \textbf{Participants trust companies they perceive to be competent and resource-rich} (Section~\ref{section:tech-companies}). In both qualitative and quantitative responses, participants indicated trust for Google over other companies.
    \item When weighing pros and cons of generic entities (e.g., government, university researchers) that might create a contact tracing app, participants go through a \textbf{complex decision process}, with no entity preferred by all and both positives and negatives about each of the entities we presented (Section~\ref{section:generic-entities}).
    \item Participants conveyed substantial \textbf{mistrust in their government} (as a generic entity) conducting contact tracing, but displayed more trust in government health agencies and, in some circumstances, judicial oversight of government (Section~\ref{section:gov-agencies-and-judges}).
\end{itemize}

\subsubsection{Participants prefer large, known, already-trusted tech companies over other tech companies}
\label{section:tech-companies}

\begin{figure}[t]
    \includegraphics[scale=.6]{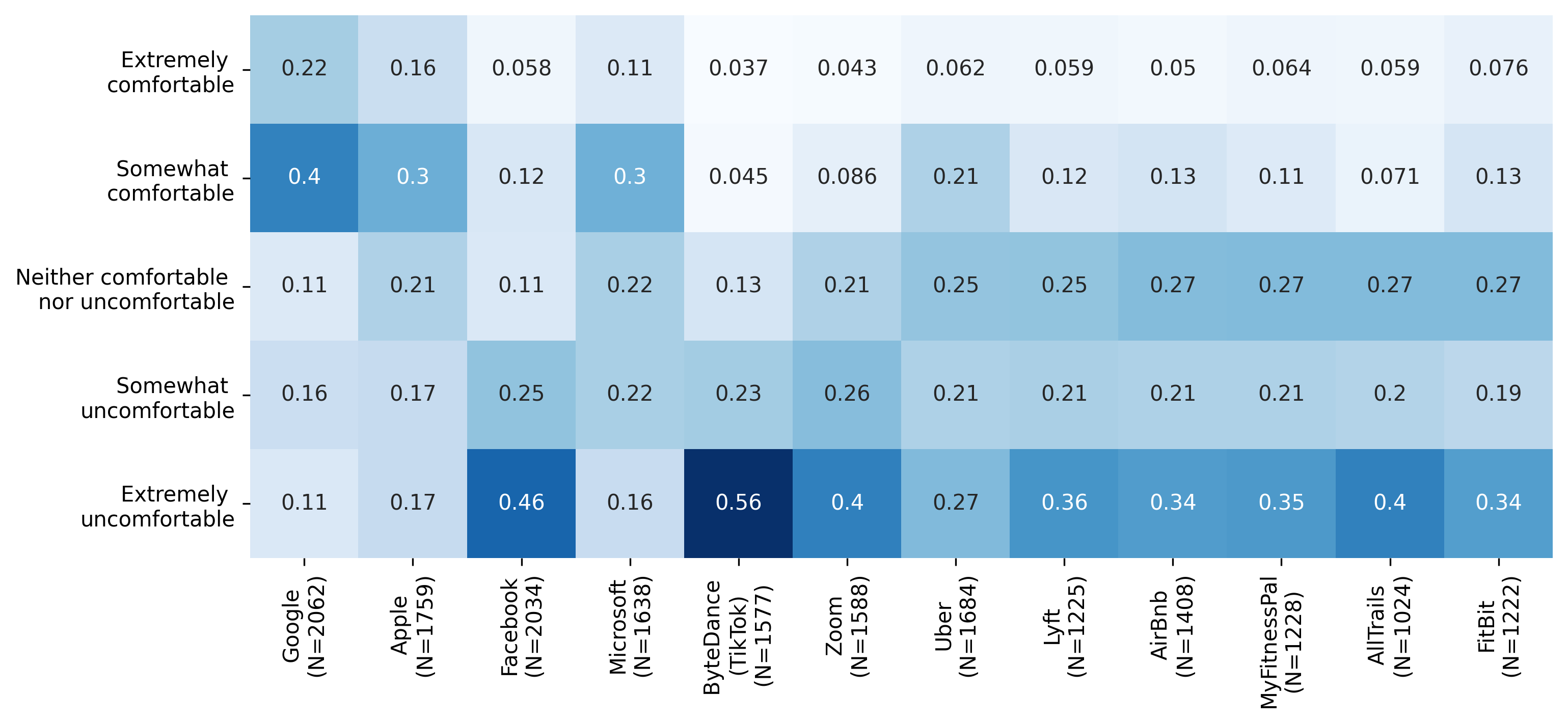}
    \caption{Participant comfort with a known company creating a \textit{new} app using their location to study or mitigate COVID-19 (Q56). Due to longitudinal stability of the data, we combine all data here in order to show nuances in opinion. Columns sum to one and represent only participants who responded to the question and did not choose the option ``I do not know enough about this company to make a decision.''}
    \label{fig:companies}
\end{figure}

Our survey asked both about a known company adding contract tracing to an \textit{existing} app vs creating a \textit{new} app to trace contacts. Though these situations require subtly different threat modeling, the results were similar; here, we present numbers from the question about a \textit{new} app since that more closely reflects today's reality. 

Participants indicated trust in large tech companies that have a good reputation concerning security and privacy and that they perceive to be capable of conducting contact tracing. More participants indicated comfort with Google or Google products than with any other company ($\mu$ = 62\% comfortable), as shown in Figure~\ref{fig:companies}. Participants also indicated some comfort with Microsoft and Apple ($\mu$ = 41 \% and 46\% comfortable, respectively), as shown in Figure~\ref{fig:companies}. Less than 30\% of participants indicated comfort with all other companies. Participants displayed the least comfort with ByteDance (TikTok) ($\mu$ = 79\% uncomfortable) and Facebook ($\mu$ = 71\% uncomfortable).

Qualitative analysis results reveals themes around user values and concerns regarding what apps they would trust most (Q23) and least (Q24) to use their location data for COVID-19 tracking. In line with the quantitative results reported above, 1205 users picked Google Maps or another Google app as their most trusted app and 431 picked Facebook as their least trusted app in the context of using location data for COVID-19 tracking. Reasons for picking their most trusted app, or for not picking their least trusted app, include the following:

\textbf{Pre-existing technical capabilities, user base, and resources.} Participants value a company that already has a large user base, sufficient monetary resources to add contact tracing to its capabilities, and the technical resources to implement accurate contact tracing (from participants' perspectives). W2P60 wrote: ``\textit{I would trust Google maps because it shows the most accurate current location real-time. I believe Google maps has the resources and manpower to allocate where I've been and when, I trust a more accurate and informational app. I imagine an app such as Instagram would be inaccurate because anyone can pick a location when they post a picture}.''

Some participants preferred an app in which location tracking is already central to its purposes (e.g., Google Maps, fitness apps, Uber, Snapchat, Pokemon Go), which they believed was a sign of technical capability to conduct contact tracing via location tracking. W8P77 explained, ``\textit{I would trust Google Maps or Apple Maps the most. Mainly because the app is made to track your location. Other apps for things like social media don't necessarily consider location as a huge factor of the app so I would feel more comfortable using an app that already tracks your location to determining spread of COVID-19.}'' Others trusted a mapping app additionally because it was \textit{not} social media and therefore maintained a degree of anonymity: \textit{``Google Maps because it does get the location access, but not more of my personal data like instagram or facebook etc, where all my contacts and photos are''} (W2P151).

\textbf{Already use and trust the app.} Participants value an app that they already know and trust either because they feel like they understand the app's data sharing and privacy procedures or because they have already ceded privacy to that app: ``\textit{Waze, it uses my data anyway so why not}'' (W1P125), and W3P3: ``\textit{I am ... locked in Apple's ecosystem, so they likely have all the data about me anyways.}'' W4P17 clarified that it was less about trust and more about risk management: ``\textit{not necesarily (sic) trust, but resignation- I know Google and Waze already know my whereabouts and am resigned to them having my data.}''

\textbf{Positive history and reputation with respect to security and privacy.} Participants preferred a company known for protecting user data and making secure and private apps. W14P85 wrote: ``\textit{I would probably trust google maps the most since most of the other apps are known to be susceptible to data breaches/leaks in the past.}'' W16P93 commented: ``\textit{I trust banking applications the most, because storing money is a serious matter,}'' while W14P43 wrote that they trusted WhatsApp the most ``\textit{as I know it’s encrypted and is very hard for mallicious (sic) hackers to break into to find my data.}'' On the other hand, W1P75 wrote that they would not trust Facebook because ``\textit{Facebook is notorious for selling user data to third parties, and I would be very uncomfortable to know that they are tracking my location with the purpose of researching COVID-19.}'' 14 participants wrote that they would not trust an app developed in China, e.g., TikTok: ``\textit{TikTok because I heard it sends user data to China}'' (W2P140). 

Participants also considered privacy policies: W5P91 preferred Apple because ``\textit{Apple have a reliable privacy policy and therefore I would trust this the most as I don’t believe any of my data would become public without my permission.}''

\textbf{Company-agnostic concerns about privacy and personal harm.} Participants also mentioned concerns that extend to any contact tracing program and reflect broader themes throughout this survey: (a) stalking or personal harm due to poorly anonymized data, (b) data leakage or privacy breaches, (c) data being sold by the company,  and (d) a ``slippery slope,'' in which this sort of tracking eventually becomes the norm.

\subsubsection{Mixed trust levels for non-corporate generic entities} \label{section:generic-entities}

\begin{figure}
    \includegraphics[scale=.6]{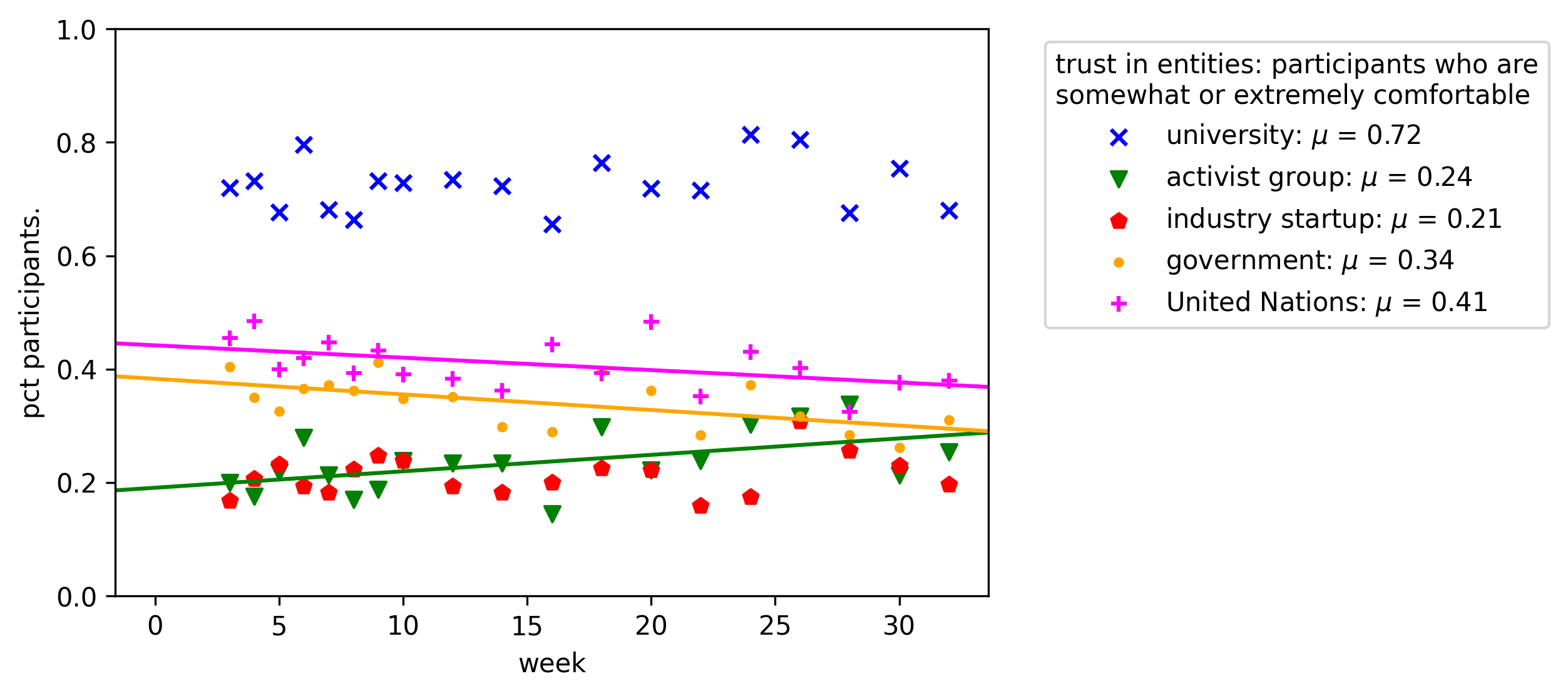}
    \caption{Participants' trust in generic entities (Q126). The higher reported trust in universities could be due to response or selection bias, as participants were shown the logo of our university before beginning the survey.}
    \label{fig:entity}
\end{figure}

Stepping back, we asked for participants' comfort with \textit{other} types of entities developing a contact tracing app. Participants indicated general mistrust for a potential new COVID-19-tracking app created by an industry startup ($\mu$ = 21\% comfortable) or an activist group ($\mu$ = 24\% comfortable), as shown in Figure~\ref{fig:entity}, but were largely split on generic trust for a government- or United Nations-developed app. Responses indicate that most would place trust in a university-developed app (72\% comfortable), but we note that at the beginning of the survey, participants were shown our university's logo and told that this survey was an academic endeavour, which may have caused response or selection bias~\cite{dell2012yours}. In contrast, Hargattai and Redmiles found that universities would be one of the least trusted entities, at less than 10\%, while a survey by the Washington Post and the University of Maryland found they were relatively trusted, at 57\%. This discrepancy highlights the need for multiple surveys and qualitative data to better understand the nuances of public opinion. 

Qualitative data revealed nuanced decisions around trust of a company or entity, echoing themes of general reputation and ability to both technically conduct contact tracing and protect data, while adding in participants' beliefs about the \textit{intentions} of a given entity. As such, 86 wrote that they would trust an app developed by scientists, universities, or researchers over any other entity.

Participants argued both for and against the entities in Figure~\ref{fig:entity}, revealing complex and individual decisions. Generally, participants indicated that trust depended on an entity's (a) intent to share or sell data, (b) anonymity or privacy guarantees, (c) reputation with respect to privacy and security, and (d) commitment to transparency and consent. Some expressed a desire for a regulatory body or for open source apps.

Some would support a tech startup because there is ``\textit{less notoriety attached to the brand}'' (W9P55) and because they do not already have \textit{other} data about the users; others would trust a big company because of its resources, credibility, and stability. Some participants considered activists unstable, unreliable, not credible, and incapable of actually securing data properly, while others valued activists groups' purer intentions, i.e., they believed that activists groups, unlike tech companies, would not sell the data on principle.

Participants who wrote about the UN mentioned its power and resources, but its international status was a plus for some (due to mistrust of their own government and not believing the UN would sell their data) and a minus for others (who disagreed with past UN work or believed the UN would be unable to produce a solution that worked for every country). W6P74 wrote in favor of the UN: ``\textit{If an International Organization such as the United Nations built the app and manages it, I will be more comfortable using the app because is a superior force than a government whom can use my data for electoral purposes or a company whom can use my data for profit.}''

Participants also raised several \textit{positives} that they would expect from a government-developed app compared to other entities: governments cannot profit off the data, can keep companies in check through policy, and have a degree of legitimacy. W8P50 wrote in support of technical and regulatory transparency: ``\textit{I would feel most comfortable if the app was open sourced for the public to be able to scrutinize, by a government agency to remove any profit motivation to misuse the data, and would feel most comfortable if the data was stored in aggregate rather than individual tracking (no data as to where I personally am at a certain time, but rather on a population level what \% are at home, near other app users, etc).}''

\subsubsection{Trust in government health agencies; mistrust for ``the government'' as a whole, with strong concerns about proper data use and sharing}\label{section:gov-agencies-and-judges}

We find that participants are more comfortable sharing location or proximity data with governmental health agencies (as opposed to other sectors of government), and that more participants are comfortable with their location or proximity data being shared with their government only if they test positive for COVID-19, echoing trends from Section~\ref{section:likelihood} and reemphasizing the need to protect the privacy of those who test positive. 

We find no strong regional trends concerning trust in government, as noted in Section~\ref{section:demographic-trends}, but other surveys have addressed this question more thoroughly and found some trends differing per country, e.g.,~\cite{altmann2020acceptability, garrett2020acceptability, kostka2020times}

\begin{figure}
    \includegraphics[scale=.6]{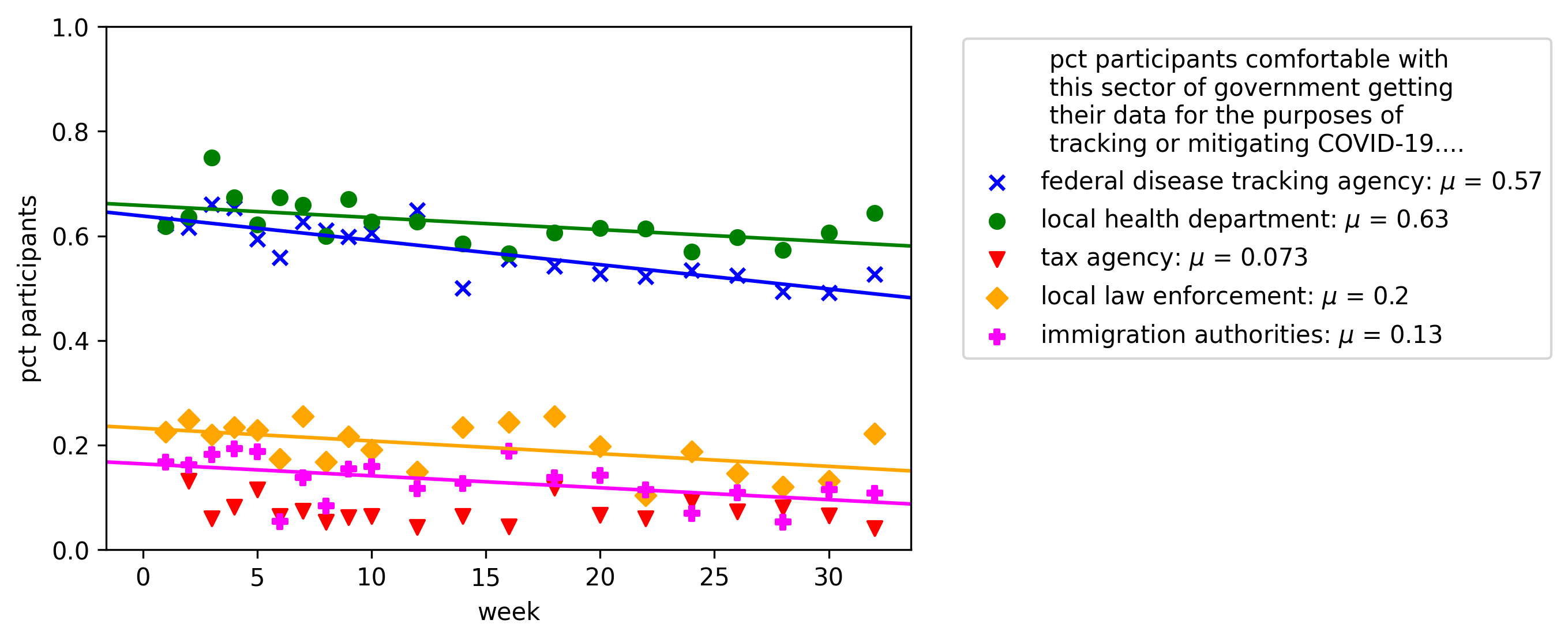}
    \caption{Participants' comfort with specific government agencies receiving their location or proximity data for the purposes of contact tracing (Q69).}
    \label{fig:gov-sectors}
\end{figure}

Participants indicated significantly higher comfort with their data being shared with health agencies for contact tracing than with other government agencies (in line with previous work~\cite{redmiles2019well}), as shown in Figure~\ref{fig:gov-sectors}. More participants were comfortable with both federal and local health departments ($\mu$ = 57\% federal; $\mu$ = 63\% local). Many fewer participants were comfortable with other agencies, i.e., local law enforcement ($\mu$ = 20\%), immigration authorities ($\mu$ = 14\%), and a tax agency ($\mu$ = 7.3\%).

\textbf{Concerns about data overuse and data sharing.} Participants indicated substantial concerns about their government's use of data and about non-consensual data sharing with or within their government. 65 participants believed that the benefits of sharing location or proximity data with their government, especially with local or national health departments, would outweigh any negatives. Of those, 39 imagined restrictions on governmental data use and retention, such as that their government should delete the data after the pandemic. 36 wrote that sharing should be voluntary, and 27 said the US government should not share the data with Immigration and Customs Enforcement (ICE). However, quantitative data indicates a lack of trust that their government would use its citizens' location data conservatively: 72\% responded that it was unlikely that their government would delete the data (Q66), and 69\% said it was unlikely that their government would use it only for COVID-19 tracking (Q67).

Participants also indicated concern that such data sharing or collection would be harmful to their safety or the safety of those in their community (Q68), with 65\% responding that they were extremely or somewhat concerned. W6P58 wrote: ``\textit{There is no chance they're going to only use it for Covid, especially in the states, and it could be very dangerous for many people, especially marginalized groups.}''

\begin{figure}
    \includegraphics[scale=.6]{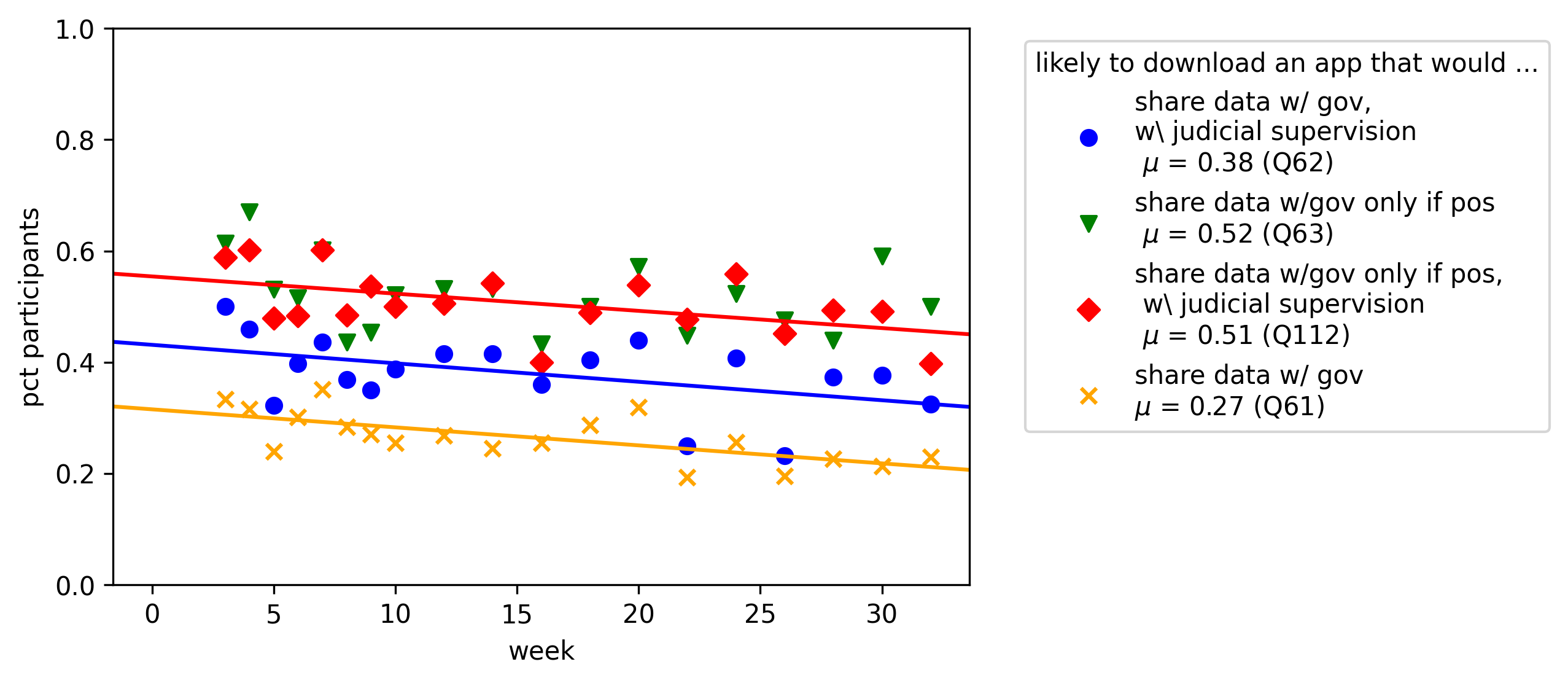}
    \caption{Participants' self-reported likelihood to download an app that shares their location data with their government under various conditions. This plot shows only those who have \textit{not} already downloaded a contact tracing app.}
    \label{fig:judiciary-supervision}
\end{figure}

\textbf{Mixed trust for judicial oversight of data sharing and use.} We observe a clear preference for judicial supervision if data is shared from users regardless of health status, and no preference if data is shared only from COVID-19 positive users, as shown in Figure~\ref{fig:judiciary-supervision}. We also observe that participants' perceptions of judicial oversight is grounded in their mental model of their judicial system and government; thus, overestimations of the level of corruption or self interest in the judiciary could skew trust and affect decision making.

Despite this preference for judicial supervision, and the increase in willingness to share if positive, 220 participants were overwhelmingly negative about judicial supervision in qualitative data, citing general concerns about not trusting their government, concerns about data sharing or usage for another purpose (142), as well as concerns about judicial impartiality (17) and tech literacy of judges (7). W30P50 (UK) wrote: ``\textit{No, judges can often be influenced by and working in a corrupt way with the government},'' and W28P76 (Poland) brought up bias and harm that could be introduced or exacerbated: ``\textit{I'm a member of a minority that our government doesn't like at this moment. I am extremely wary.}'' Other participants felt negatively because they did not believe the judges would make the right decisions: ``\textit{The judges here in Canada suck, and can't be trusted to deal out justice properly.  We have a revolving door justice system for criminals - what makes you think they'd do any better when things aren't as clear cut as criminal cases?}'' (W22P88).

Participants commented on judges' digital literacy, citing that ``\textit{government officials, globally, seem to have a high rate of technology illiteracy}'' (W14P83, Ireland). Instead, one participant suggested that there be ``\textit{data watchdogs and possibly even a human rights person}'' (W18P98, Slovenia). 

Some participants who were concerned about judicial impartiality actually desired oversight, but thought it would not be possible in their country due to corrupt or politically motivated judges: ``\textit{Would be more influenced if observed by an independent party not affiliated with the government or partners}''(W24P89, UK). W22P55 (US) wrote: ``\textit{Judicial oversight is a good starting point. But with the current administration, I feel like trust in the judicial system has been slowly getting eroded.}'' Alluding to different levels of trust, understanding, and different political systems, W26P89 (Chile) wrote that``\textit{Judges in my country are not really that much better than politicians....}''

Despite many concerns about corruption, politicization, and bias, some participants did present positive values, including already trusting the judicial system and/or their government (115) or judicial oversight being better than nothing (24). W28P11 (UK) felt that judicial oversight would prevent corruption, instead of enabling it; they wrote that judicial oversight ``\textit{should prevent abuse of power}.'' W26P71 (Portugal) wrote: ``\textit{I think judicial oversight of apps should be more common}.'' Additionally, themes of consent arose (27) along with the idea that such oversight is already occurring (21). Echoing themes from Section~\ref{section:likelihood}, 32 participants reiterated that their government should have access to their location data only if they actually tested positive for COVID-19.  

This combination of qualitative and quantitative data tells a complex story about participants' trust in their governments, including trust in the legal system, as well as their understanding of how both judicial oversight and contact tracing operate. It also reflects the seemingly directly competing values of privacy and altruism that push people towards not sharing data and pull them towards sharing it for the common good. Again, participants' mental models of the mechanism in question here, the judicial system, may be inaccurate or incomplete, but still drive their willingness to participate in automated contact tracing. Additionally, political and judicial systems differ across the globe, and judicial oversight may be appropriate in some countries and not others.

\subsection{Mixed attitudes towards alternate data sources}
\label{section:alternate-sources}

We now review participants' opinions about data sources \textit{other than} smartphone apps: cell tower location data, credit card history data, public sensors (including surveillance cameras), and electronic wearables. We find that:
\begin{itemize}[leftmargin=*]
    \item There is \textbf{more support for contact tracing using cell tower location data than for the other non-app data sources} we asked about. Additionally, about the same percentage of participants supported the most popular cell tower data question (in which location data of COVID-19-positive users would be shared with the government) as participants who said they were comfortable with proximity tracing, app data from positive COVID-19 users being shared with their government, or app data being used by the app developer but not shared further (all around 50\%). However, these comparisons must be interpreted with caution due to (1) potential biases arising from question ordering, and (2) the fact that the app questions ask about ``likelihood'' whereas the questions reported on here ask about ``comfort.''

    \item We also find that many of the concerns and values about smartphone contact tracing are magnified with non-smartphone automated contact tracing. Specifically, we observe that \textbf{consent for data collection, use, and sharing is extremely important to participants}, and particularly relevant to these non-smartphone data sources, which can largely occur without the user's informed consent. We call on the stakeholders in power to critically examine the need for consent beyond a terms of service agreement.

\end{itemize}

\subsubsection{Support for cell tower data over other data sources}

\begin{figure}[t]
\begin{subfigure}[t]{.5\textwidth}
  \centering
  \includegraphics[width=.9\linewidth]{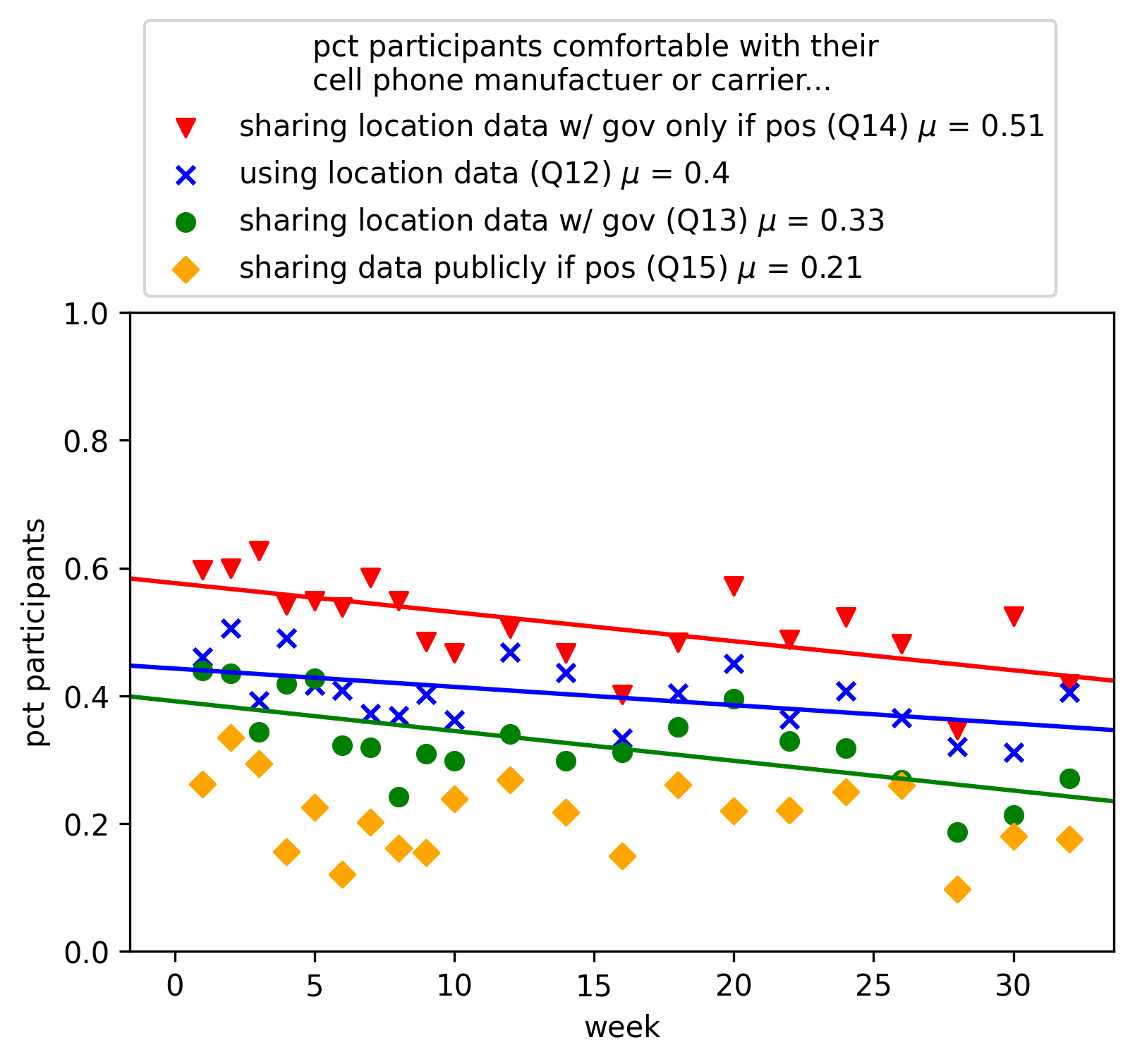}  
  \caption{Attitudes towards cell tower location data being used for COVID-19 tracking: participants who said they were somewhat or extremely comfortable with their cell phone manufacturer or carrier using their location data for the purposes of COVID-19 tracking.}
  \label{fig:cell_tower_data}
\end{subfigure}~\hspace{15pt}
\begin{subfigure}[t]{.5\textwidth}
  \centering
  \includegraphics[width=.9\linewidth]{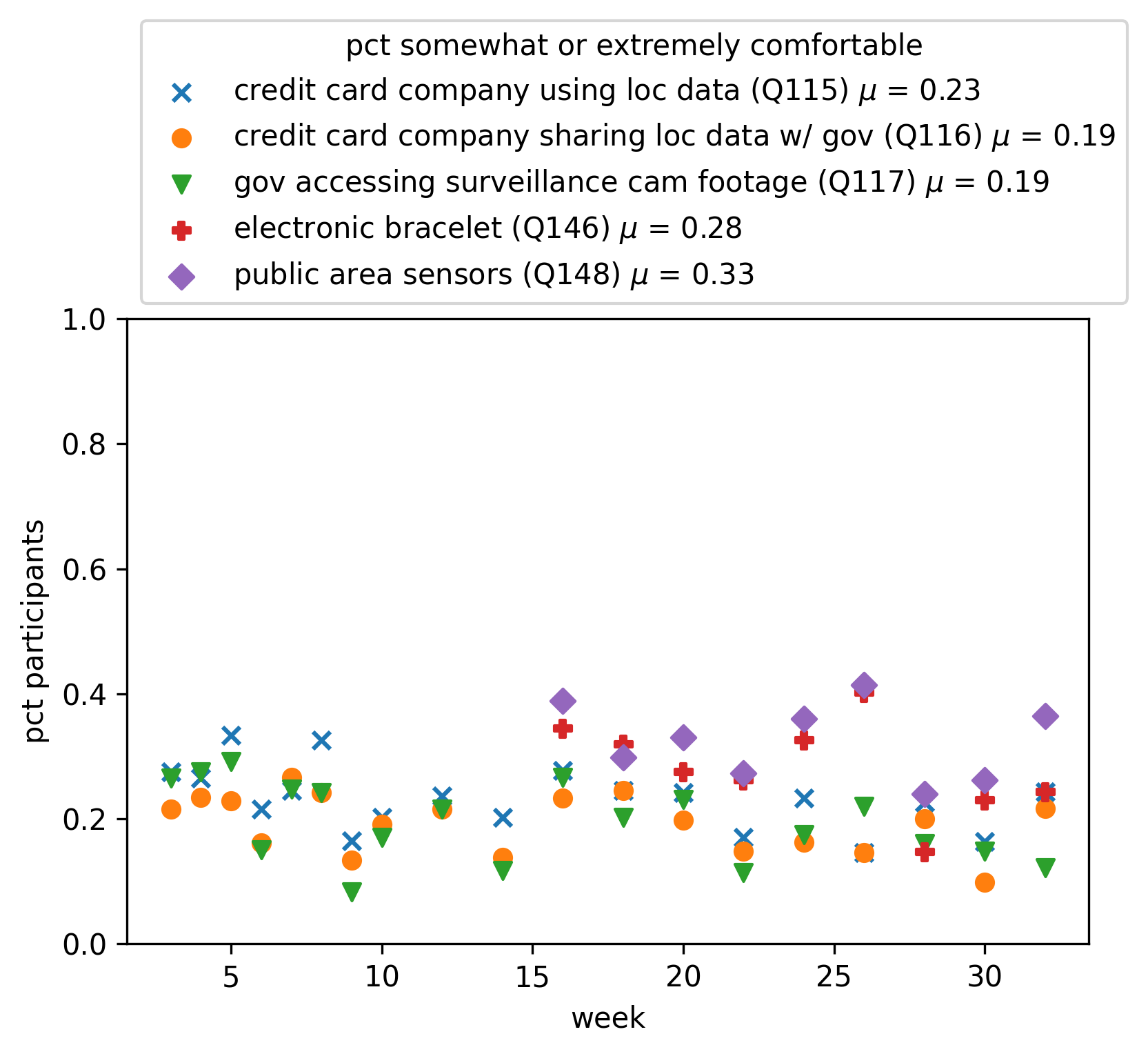}   
  \caption{Attitudes towards other data sources: participants who said they were somewhat or extremely comfortable with electronic wearables, public sensors, surveillance cameras, or credit card data being used for contact tracing.}
  \label{fig:ccs-wearables-etc}
\end{subfigure}
\caption{Participants' attitudes about cell tower data and other data sources.}
\label{fig:other-data}
\end{figure} 

Regarding comfort with cell tower data being used for contact tracing, participants were most comfortable with their government being given the data if they tested positive, and many fewer were comfortable with the data being released publicly, as shown in Figure~\ref{fig:cell_tower_data}. This preference resembles the increased comfort with data sharing \textit{if positive} (Figures~\ref{fig:judiciary-supervision} and~\ref{fig:how-likely}, Sections~\ref{section:likelihood} and~\ref{section:who-develops}). 

We observe statistically significant but slight downward slopes for three of the questions, and we note that the order of participants' comfort is largely consistent across weeks: the most participants were comfortable with their cell phone manufacturer or carrier sharing  data with their government if they are positive ($\mu$ = 51\%), and the fewest were comfortable with their location history being shared publicly if they tested positive ($\mu$ = 21\%). This sentiment of being uncomfortable with public disclosure more generally underscores the importance of contact tracing data being properly protected when and if collected to avoid data exposure were a breach to occur.

As shown in Figure~\ref{fig:other-data}, participants were much less comfortable with their credit card history being used for contact tracing: $\mu$ = 23\% comfortable with contact tracing done by the credit card company, and $\mu$ = 19\% comfortable with that data going to their government. Participants were generally uncomfortable with their government using footage from surveillance cameras or other public area sensors for contact tracing ($\mu$ = 19\% comfortable with surveillance cameras and $\mu$ = 33\% comfortable with public area sensors) as well as with electronic bracelets ($\mu$ = 28\%).

\subsubsection{Qualitative data reveals privacy concerns and the need for informed consent}

Through the qualitative data, we find similar thematic concerns as from Section~\ref{section:who-develops} about privacy, data sharing, surveillance, equity, accuracy. Participants also reiterate the need for meaningful consent and transparency. We emphasize, again, that there is no perfect data source, and that users will likely have privacy concerns about any potential source. As such, the technology and policy communities must assume responsibility for protecting and informing users about personal data acquisition and use.

\textbf{Concerns about privacy, anonymity, and data sharing regarding alternate data sources.} Privacy and anonymity were of paramount concern, especially with regard to public sensors and cell tower data, with many mentioning a ``surveillance state.'' 88 participants wrote that the use of cell tower data for contact tracing could be a ``slippery slope'' towards more permanent privacy invasion or other misuse of data. 31 specifically referenced George Orwell's 1984 in reference to contact tracing data from surveillance cameras. W3P122 wrote: ``\textit{I’m very against expanding the surveillance state, even for a good reason, because it’s never going to get rolled back.}'' 67 participants were also uncomfortable with the idea of wearable electronics, with 22 specifically associating it with feeling like a ``criminal,'' ``prisoner,'' or ``animal.''

132 participants had concerns about the privacy of cell tower data, focused on the data being publicized or shared with their government (see Figure~\ref{fig:cell_tower_data}); 68 of those specifically mentioned anonymity as a value. In responses about using credit card data, 53 specifically considered the sensitive nature of financial data. Some participants considered credit card data less private because it ``\textit{gives only a handful of specific locations, and not a complete timeline of every location like a phone would}'' (W16P59), while others mentioned a mistrust of financial institutions (``\textit{Credit card companies are not customer friendly and are always behind monetary benefits therefore I would not like them to have my data and trust them}'' W3P20). These responses also suggest that participants' mental models of the privacy of their credit card history is that location tracking would involve revealing their purchase history.

\textbf{Concerns about equity, discrimination, and personal harm  regarding alternate data sources.} Participants raised concerns about the potential for discrimination, harm, and equity, echoing concerns from privacy experts~\cite{efficacyCTapps} and emphasizing the need for technologists and policy makers to take extraordinary and thorough measures to protect potentially vulnerable populations. 17 participants feared harassment, prosecution, or discrimination with the use of cell tower data: ``\textit{if the location of people that has (sic) tested positive for COVID-19 is publicly shared, they might get targeted and hurt (or worse). This last idea comes from the fact that this was a situation given in my country, where it was publicly shared that a group of immigrants was tested positive, and this lead to them being persecuted}'' (W16P98, Chile). Another participant mentioned the concern ``\textit{that people would would draw conclusions from (for example) two people being at the same hotel at the same time}'' (W20P89). This situation is reminiscent of South Korea's initial handling of location tracking: location data and biographical details were posted publicly and were not sufficiently anonymized; groups discovered the identities of those who had tested positive and rumors started about extramarital affairs and plastic surgery trips~\cite{southKoreaRumors}. South Korea has since started anonymizing the publicly released data more thoroughly~\cite{singerAsCoronavirusNYT}. 

Speaking to a theme of equity, some participants wrote that contact tracing using credit card data would be ineffective because ``\textit{credit cards are only for the elite}'' (W3P33, South Africa). W3P63 extrapolated further, raising concerns about the potential societal implications of health information being linked to financial status: ``\textit{I fear this will lead to access to credit and my credit score being linked to my health or my compliance with social distancing. I am social distancing, but I worry about the future implications for this.}''

Participants also described the potential for harm caused through racial bias, specifically in reference to public sensors and surveillance cameras, recalling numerous issues with existing facial recognition systems. Others were concerned about the potential for future misuse: ``\textit{Police have already been caught illegally using face ID software, I certainly wouldn't trust them [using surveillance camera data for contact tracing]}'' (W7P30).

\textbf{Concerns about accuracy regarding alternate data sources.} Echoing themes from Sections~\ref{section:likelihood} and~\ref{section:who-develops}, participants reasoned about the accuracy of alternate data sources, concerned that cell tower data, credit card data, public sensors, and electronic wearables could not provide sufficient data for accurate contact tracing. 94 participants raised concerns about the lack of accuracy of credit card data: W9P60 noted that credit card data would an inappropriate source of contact tracing data because it is ``\textit{not good enough to track movement. Park/beach and many more places where I would not use card but be around people.}'' Participants also reflected on the potential for facial identification from surveillance cameras to fail (e.g., if wearing masks or sunglasses) or to be impractical due to cost or lack of population density. Reflecting on practicality, W16P71 wrote: ``\textit{A tracking bracelet may get lost or people may forget to wear it when they leave the house. It wouldn't provide the most accurate data.}''

\textbf{Participants value consent and transparency  regarding alternate data sources.} Throughout all questions, participants raised concerns about consent and transparency, which are particularly important with data that could be collected without their knowledge or with minimal consent (e.g., through a terms of service agreement), for example, via public sensors or data that already exists, like credit card data or cell phone tower data. W1P190 wrote that an end-date for cell tower data collection and use would make them feel more comfortable: ``\textit{If I were informed in advance of the initiative and there was a sunset date for the initiative, I'd be somewhat likely to allow it for the purpose of scientific research. Absent my explicit consent and for only a short period of time, however, I'd be extremely uncomfortable with it.}'' Even W12P86, who was largely comfortable with the  use of credit card data for contact tracing, gave the caveat that ``explicit consent'' was necessary: ``\textit{I feel more comfortable with my location being acquired (with my knowledge and consent) via this method as it makes me more relieved that I am not actively being tracked (or my location is not being tracked and traced to beach movement). The usage of my credit card history allows me windows of privacy which I would not want anyone interfering with}.'' Thus, policy makers and technologists must both work to protect and inform users.

\textbf{Some support for non-smartphone data sources for contact tracing.} Though a majority of participants raised concerns, some supported the alternative data sources we asked about. Of those who supported (or did not oppose) the idea of cell tower data being used for contact tracing, 123 mentioned altruism and the greater good of contact tracing, suggesting that many users, under the right circumstances, may decide that the greater good of their communities outweighs some personal privacy concerns. 24 said they would accept such data sharing if it were for research, while 20 would accept only if they tested positive for COVID-19; these opinions raise questions about these participants' mental models of contact tracing or whether they would consider such data sharing as a way to reduce their need to quarantine if positive.

\subsection{Demographic trends} \label{section:demographic-trends}

Given the lack of strong longitudinal trends in many of the questions, we examined demographic trends by combining data from \textit{all} weeks. We find no trends for age, gender, or time spent outside home. We also observe no correlation with infection rate (see Figure~\ref{fig:world-covid}). The following section presents trends that are largely present throughout all questions; we show the questions related to willingness to use a COVID-19 contact tracing app.

\begin{figure}[t]
\begin{subfigure}[t]{.5\textwidth}
  \centering
  \includegraphics[width=.9\linewidth]{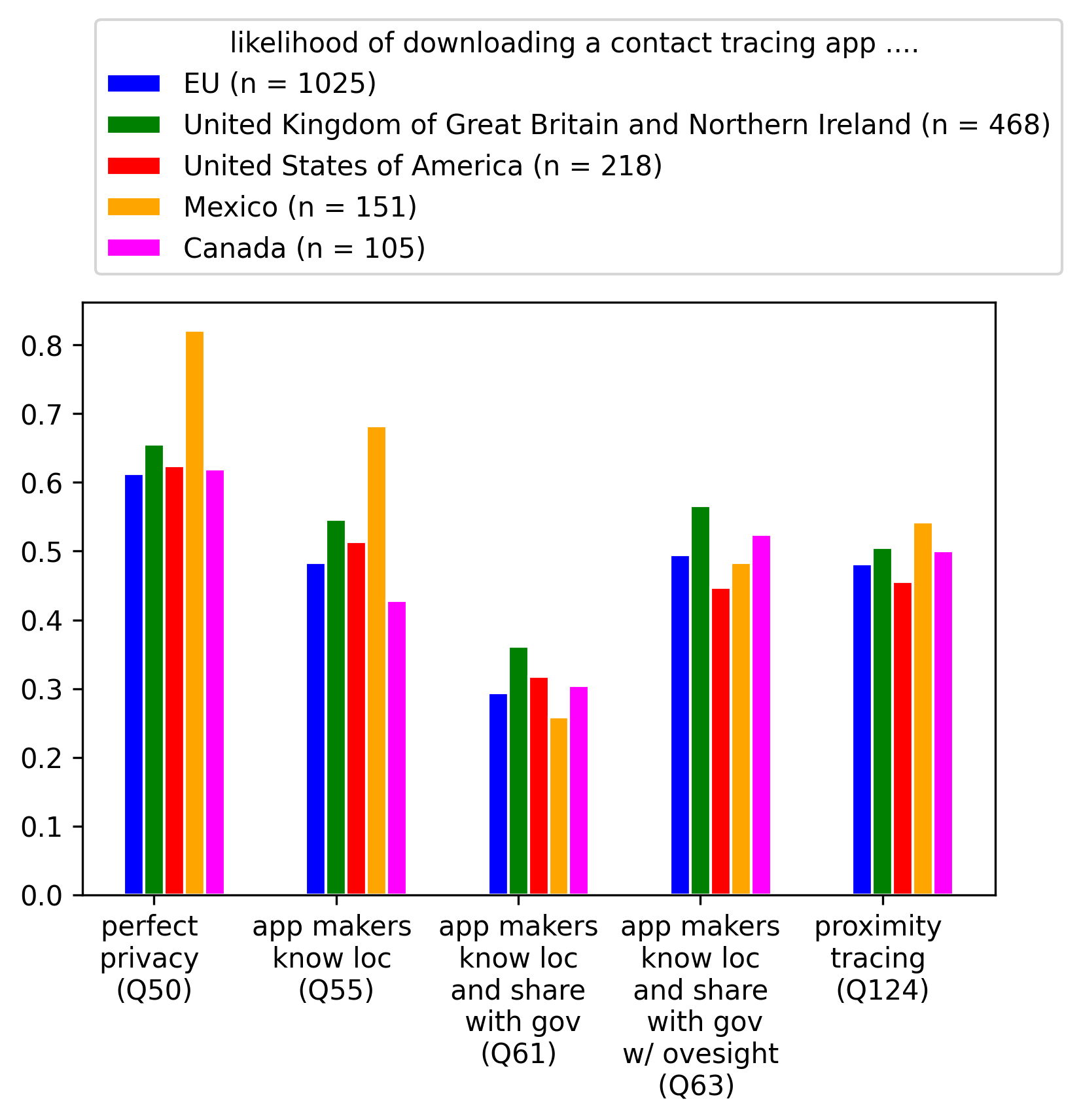}  
  \caption{Likelihood of downloading a contact tracing app versus location (for the countries from which we had at least 100 participants total). (See Section~\ref{section:likelihood}.) }
  \label{fig:regional-trends-willingness-to-download}
\end{subfigure}~\hspace{15pt}
\begin{subfigure}[t]{.5\textwidth}
  \centering
  \includegraphics[width=.9\linewidth]{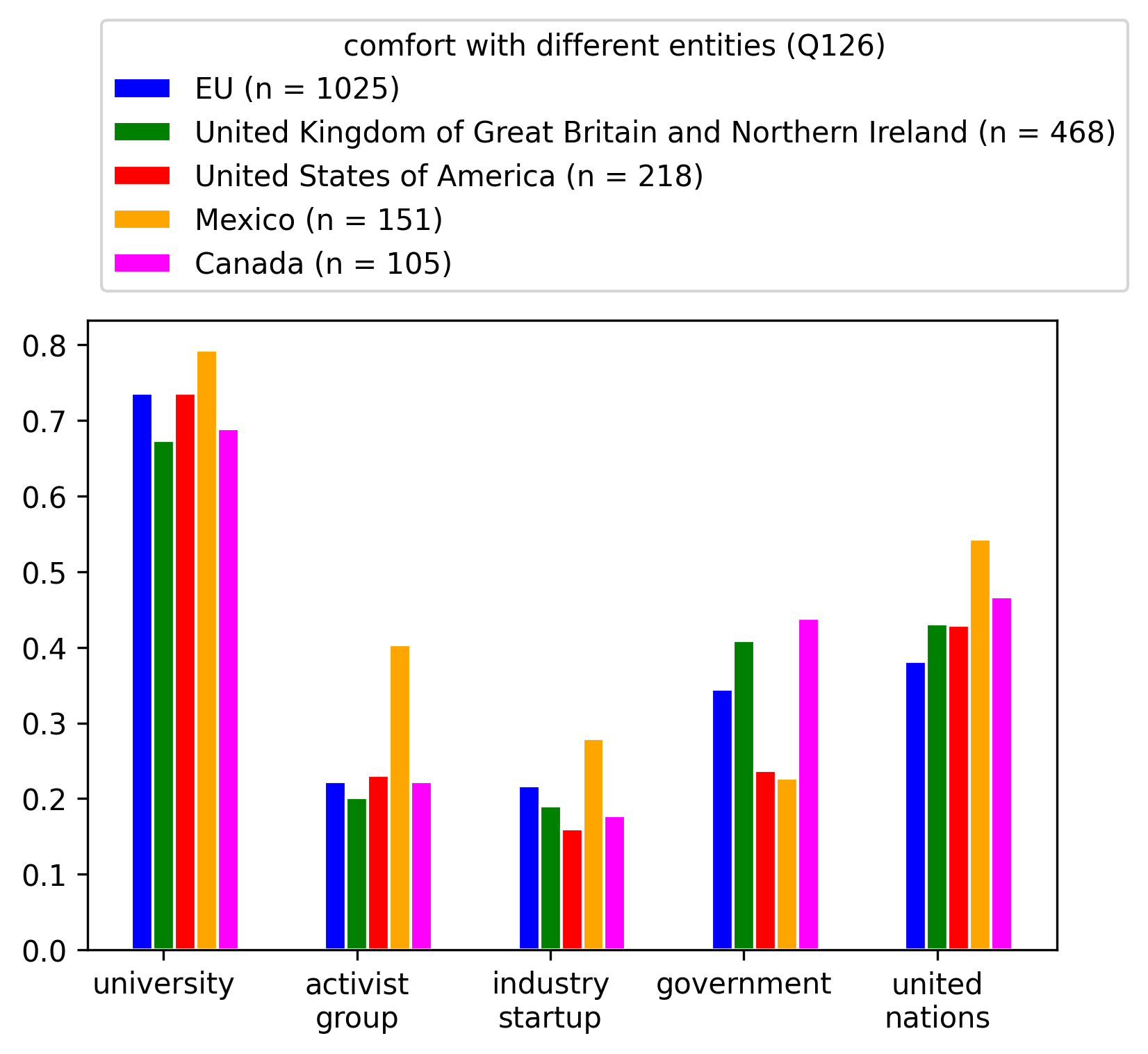}   
  \caption{Comfort with generic entities creating a contact tracing app (for the countries for which we had at least 100 participants). (See Section~\ref{section:generic-entities}.)}
  \label{fig:regional-trends-gov-sectors}
\end{subfigure}
\caption{These figures show two sets of questions broken down by regional responses.}
\label{fig:regional-trends}
\end{figure}

\textbf{Few regional trends appear in our dataset, but related work investigates them more thoroughly.} When examining regional trends, we include only regions (e.g., the EU) or countries from which we had more than 100 participants: EU, UK, USA, Mexico, and Canada. We do not find regional trends between the EU, UK, USA and Canada. However, we find that participants from Mexico are less willing to share data with their government, but perhaps more willing to give up privacy if their government is not involved, as shown in Figure~\ref{fig:regional-trends} (as compared to participants from the EU, UK, US, and Canada). Others have studied regional differences in attitudes towards contact tracing applications~\cite{altmann2020acceptability, abeler2020support, kostka2020times}, and there is a growing body work about cultural or regional differences in privacy attitudes and definitions more generally (e.g.,~\cite{sawaya2017self, harbach2016keep}).

\begin{figure}[t]
\begin{subfigure}[t]{.5\textwidth}
  \centering
  \includegraphics[width=.9\linewidth]{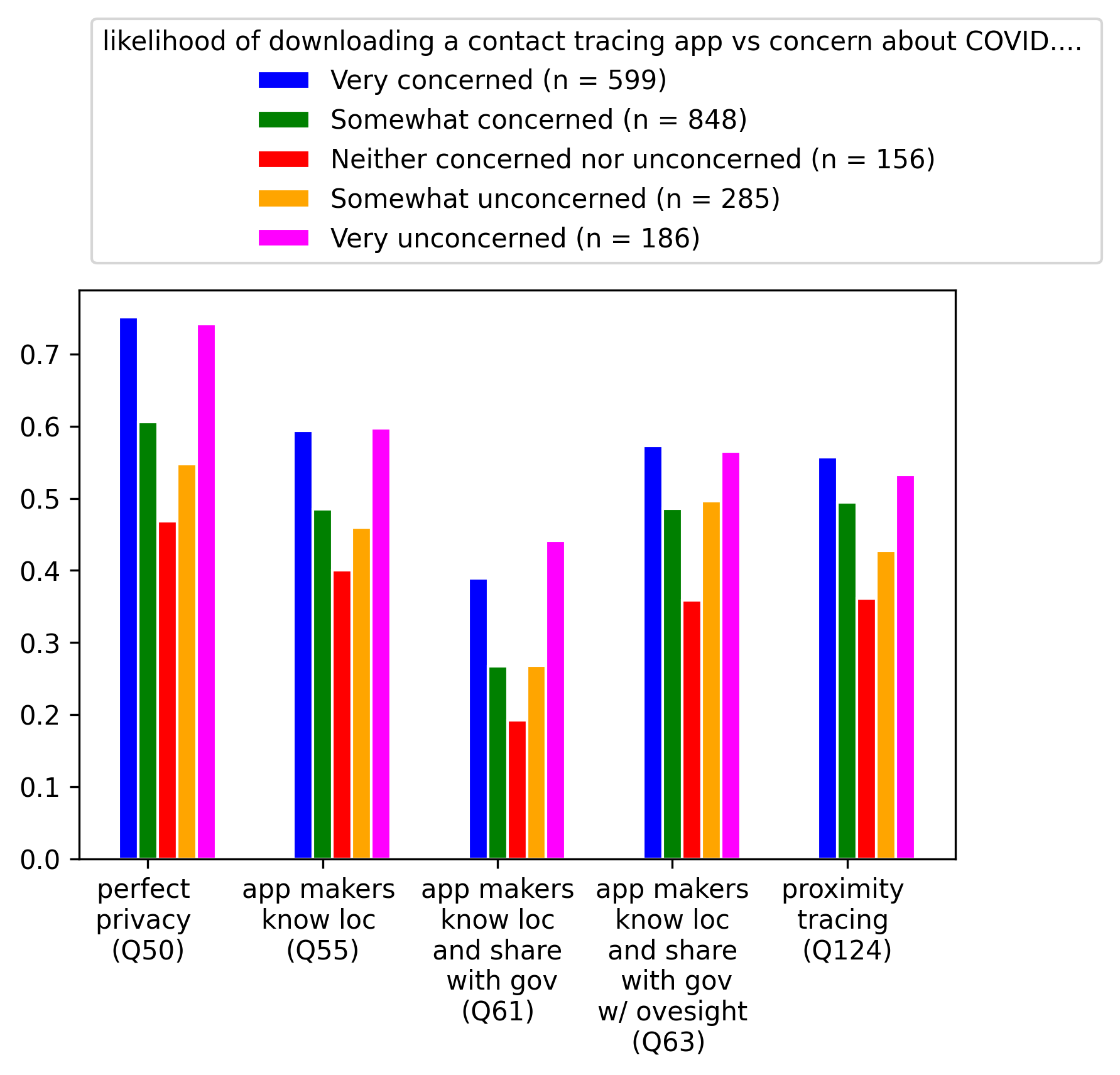}  
  \caption{}
  \label{fig:covid-concern-willingness-to-dl}
\end{subfigure}~\hspace{15pt}
\begin{subfigure}[t]{.5\textwidth}
  \centering
  \includegraphics[width=.9\linewidth]{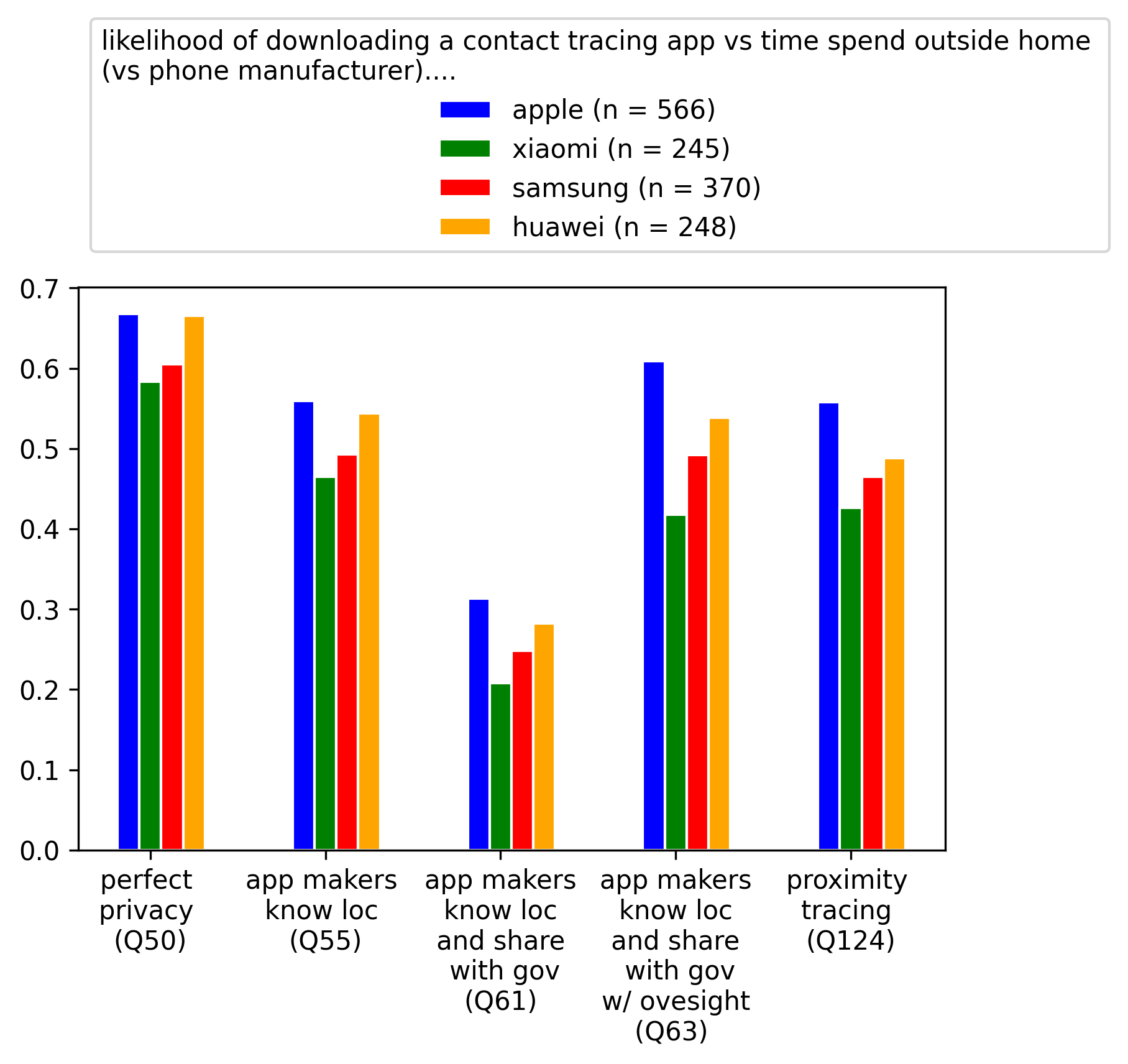}   
  \caption{}
  \label{fig:phone-willingness-to-dl}
\end{subfigure}
\caption{Willingness to download as compared to (a) concern about COVID-19 and (b) phone manufacturer.}
\label{fig:demographic-willingness-to-dl}
\end{figure}

\textbf{Concern (or lack of concern) about COVID-19 correlated to willingness to download a contact tracing app or give up privacy.} Perhaps unsurprisingly, as shown in Figure~\ref{fig:covid-concern-willingness-to-dl}, we find that extreme concern about COVID-19 is correlated with greater willingness to download a contact tracing app or surrender some personal privacy for the sake of contact tracing. However, we also find that extreme \textit{unconcern} is correlated with the same willingness to download a contact tracing app or give up some privacy. One possible explanation is that those who are unconcerned are more accepting of risk in general and thus may be more willing to take actions that others view as potential violations of privacy.

\textbf{Phone manufacturer correlated to willingness to download a contact tracing app or give up privacy.} As seen in Figure~\ref{fig:phone-willingness-to-dl}, we find that participants' choice of phone manufacturer correlates with their willingness to cede some privacy, with Apple users being the most willing, followed by users of Huawei, Samsung, and Xiaomi phones. This willingness has multiple possible explanations, including trust in the phone manufacturer itself, and may run along other demographic lines that we are not aware of.

\section{Discussion} \label{section:discussion}

Drawing from our findings, we surface lessons for both researchers and stakeholders (including app makers, public health experts, policy makers, and others). 

\textbf{User education is needed to correct inaccurate mental models and therefore enable adoption.} Users are concerned with the accuracy of the technology involved in contact tracing as well as companies' abilities to actually conduct contact tracing, but may be ill-equipped to accurately reason about these factors due to an understandable lack of technical training. Adding to recommendations by groups with similar findings~\cite{zhang2020americans, williams2020public}, we recommend that technology companies and governments conduct user education campaigns to teach users---at an appropriate technical level---to reason about the extent to which the contact tracing app available to them is appropriate for their personal situation.
    
\textbf{Users value transparency and consent, and may be less concerned about privacy if they feel in control.} Our data revealed substantial fears about data overuse and oversharing, echoing privacy concerns found by numerous others (see Section~\ref{section:related-work}). Participants feared non-consensual data sharing with both their own government and foreign governments as well as data being used for purposes other than contact tracing (e.g., advertising, national security, etc). We recommend that policy makers continue to both create restrictive policies to make users comfortable and educate users about those policies.

\textbf{An individual's willingness to download a contact tracing app depends on security and privacy \textit{and other factors.}} Beyond the privacy and security concerns and opinions that our work surfaces, there are many other broader issues that must be addressed with the release of a contact tracing application, some of which arose in our qualitative data: participants brought up concerns of accuracy, equity, and access to smartphones, as well as concerns about the harms of data overusage or sharing disproportionately affecting certain parts of the population. Building on those themes, we encourage stakeholders to consider \textit{accessibility and usability by all}, including those who do not speak the majority language, those who cannot read or write, those who have one or more disabilities (e.g. vision impairment). Additionally, \textit{not all people have smartphones}, and some high risk groups, such as seniors, may be less likely to regularly use a smartphone. A smartphone app excludes those sets of users. If a certain demographic group is left without access, or without usable access, they will benefit less from contact tracing, potentially resulting in different rates of infection. Thus, we urge stakeholders to explore acceptance for automated contact tracing in a broader context than strictly security and privacy and refer readers to~\cite{equityContactTracing, equityContactTracing2, efficacyCTapps, effCivilLivertiesPublicHealth, redmilesUserConcerns, raskarAppsGoneRogue} for a fuller discussion of equity and efficacy concerns. 

\textbf{There are substantial challenges with future-proofing longitudinal work during a rapidly evolving global event due to changing terminology and technology.} Terminology and technology have evolved \textit{rapidly} during the COVID-19 pandemic, and we thus implore other researchers doing longitudinal work to carefully consider the phrasing of their surveys. In designing our survey in late March, we knew we were trying to design questions that would remain relevant for months, throughout a rapidly changing world event. We designed our survey with a goal of being resilient to such world changes. We include our full survey in the appendix, and explicitly note when new questions were added, as the world and the global discussion around contact tracing evolved. Our key recommendations, to any future designers of surveys focused on rapidly evolving issues, are to: (1) design the initial survey with an eye toward future-resiliency,  (2) strive to make sure that any additions or modifications to the survey do not invalidate longitudinal analyses, and (3) clearly document any such changes, so that the scientific community can fully evaluate the work and the results.

\textbf{Researchers should continue to study acceptance for automated contact tracing within \textit{specific} populations.} Our survey focused on a longitudinal view of young, white, European and North American views towards automated contact tracing, but we were unable to study any one particular population in depth. Other work has studied populations at a single-country level, e.g., the Netherlands, Germany, Australia, but to our knowledge, few have focused yet on specific and potentially vulnerable subpopulations or minorities, who might have heightened or different privacy preferences, and who also might have greater vulnerability to the virus (one exception is Filer et al., who studied adoption and attitudes amongst health care workers in England's national health care system, the NHS~\cite{Filer2020test}). We specifically call for further research on minority populations that may be harder hit by the virus (e.g., communities of color in the US~\cite{kabarriti2020association}), communities that may have a more strained relationship with government or authorities (e.g. Black communities in the US, undocumented immigrants, political dissidents), or communities that may remember a past epidemic (e.g., gay men who lived through the AIDS epidemic). Despite certain communities being particularly vulnerable, we are not aware of existing studies about contact tracing and privacy for such populations, and we believe it is crucial for future work to study and address the specific contexts of these groups.

\section{Conclusion}

Here we have presented results from a longitudinal survey about public opinion surrounding location privacy and contact tracing during the COVID-19 pandemic, finding that public opinion is largely stable over time, and that they have significant and diverse privacy concerns about contact tracing.

The report adds to other concurrent work about public opinion on potential contact tracing technologies and privacy concerns, and we strongly encourage contact tracing developers, policy makers, and others to consider the user values and concerns presented here, as user cooperation is crucial.

\section{Acknowledgements}


We are very thankful to Karl Weintraub for general discussions, sanity, and his Pandas knowledge. We are also thankful to Gennie Gebhart for early discussions about this work. We thank Christine Chen, Ivan Evtimov, Joseph Jaeger, Shrirang Mare, Alison Simko, Robert Simko, and Eric Zeng for their valuable feedback on pilot versions of our survey. Additionally, we are thankful to Gennie Gebhart, Joseph Jaeger, and Elissa Redmiles for their valuable feedback on a draft of this paper. We also thank Alison Simko and Sandy Kaplan for their brilliant editing, and Lang Liu, Kristof Glauninger, and Zhitao Yu from the University of Washington's Statistics Consulting program.

We are also very thankful to Prolific for supporting our research by waiving their fees for one month as part of their COVID-19 fee waiver program. 






This research was supported in part by a University of Washington Population Health Initiative's COVID-19 Rapid Response Grant and by the University of Washington Tech Policy Lab, which receives support from: the William and Flora Hewlett Foundation, the John D. and Catherine T. MacArthur Foundation, Microsoft, the Pierre and Pamela Omidyar Fund at the Silicon Valley Community Foundation. This work was also supported by the US National Science Foundation (Awards 1565252 and 1513584). 



\bibliographystyle{ACM-Reference-Format}
\bibliography{bibliography}

\appendix
\appendix
\section{Survey Protocol}
\label{survey-protocol}
The survey protocol is below. We give section headings and descriptors for the reader's reference here; participants did not see headers. Unless otherwise specified, all questions were answered on a five-point Likert scale.

The logo of our institution (with the institution name prominent) appear as a header to each survey page. Our lab and department name did not.

\subsection{Consent and Screening}

This is a survey about \textbf{location tracking and Coronavirus (COVID-19)} by researchers at the University of Washington, in Seattle, Washington, USA. The University of Washington’s Human Subjects Division reviewed our study, and determined that it was exempt from federal human subjects regulation. We do not expect that this survey will put you at any risk for harm, and you don’t have to answer any question that makes you uncomfortable. In order to participate, you must be at least 18 years old, regularly use a smartphone, and able to complete the survey in English. We expect this survey will take about 15-20 minutes to complete.

If you have any questions about this survey, you may email us at <study-specific-email>.

Thanks for taking our survey! To start, please answer the two questions below...

Are you at least 18 years old? [yes, no]

Do you use a smartphone regularly? [yes, no]

\subsection{Demographics I}

This survey involves questions about \textbf{COVID-19}, the disease caused by SARS-CoV-2 (commonly known as \textbf{coronavirus}).

Q6: How concerned are you about COVID-19? 

Q7: Do you believe that social distancing is an important tool for slowing the spread of COVID-19? [yes, no, not sure]

Q8: Averaged over the past week, approximately how many hours much time per day did you spend out of your home, within 6 feet (2 meters) of other people? (e.g., getting groceries, working at an essential job like in a hospital, in a grocery store, etc). 
[`I did not leave my home', 0-1 hours per day, 2-3 hours per day, ... , 7-8 hours per day, 8+ hours per day] 

Q144: Do you believe that wearing a mask is an important tool for slowing COVID-19? [yes, no, not sure]

Q145: Over the past week, how often did you wear a mask when you were out of your home? [All of the time, most of the time, some of the time, rarely, never]

Q9: In which country do you currently reside? [drop-down country list]

Q10: For respondents in the USA: in which state do you currently reside? [drop-down US state list]

\subsection{Cell phone manufacturer and provider location data}

\textit{Cell phone manufacturers and cellular providers have access to your physical-world location.}

Q12: How comfortable are you with your cell phone manufacturer or your cellular carrier using your location data for the purposes of studying or mitigating the spread of COVID-19? 

Q13: How comfortable are you with your cell phone manufacturer or your cellular carrier sharing your location data for the past two weeks \textbf{with your government} for the purposes of studying or mitigating the spread of COVID-19? (\textbf{Regardless of whether you test positive for COVID-19.})

Q14: \textbf{If you tested positive for COVID-19}, how comfortable would you be with your cell phone manufacturer or your cellular carrier sharing your location data for the past two weeks \textbf{with your government} for the purposes of studying or mitigating the spread of COVID-19?

Q15: \textbf{If you tested positive for COVID-19}, how comfortable would you be with your cell phone manufacturer or your cellular carrier sharing your location data for the past two weeks \textbf{publicly}?

Q16: Optionally, do you have any other thoughts about your cell phone manufacturer or your cellular carrier sharing your location data for the purposes of studying or mitigating the spread of COVID-19? [free response]

\subsection{Existing app location data}
\textit{Some phone applications have access to your physical-world location, either when the application is in use or all the time. \textbf{Suppose the makers of an existing app on your phone started using your GPS location data to study or mitigate the spread of COVID-19.} For example, this could include disclosing past locations of known positive COVID-19 cases to the public or to the government, or alerting people who have crossed paths with the positive case.}

Q18: Below we've listed 15 commonly-used apps. For the apps that you use regularly: how comfortable are you with the following apps using your location data for the purposes of studying or mitigating the spread of COVID-19? [``I don’t use this app'' + 5-point Likert scale for each of the following apps]

\begin{itemize}
    \item Google Maps
    \item Apple Maps
    \item Waze
    \item Facebook
    \item Instagram
    \item TikTok
    \item WhatsApp
    \item Facebook Messenger
    \item Zoom
    \item Uber
    \item Lyft
    \item Airbnb
    \item Calorie Counter (MyFitnessPal)
    \item FitBit
    \item AllTrails
\end{itemize}

\textit{Suppose that one of the apps that you regularly use  -- not necessarily one of the ones above -- started using your location data to study or mitigate the spread of COVID-19. }

Q20: How comfortable are you with this app using your location data for the purposes of studying or mitigating the spread of COVID-19? 

Q22: \textbf{If you tested positive for COVID-19}, how comfortable would you be with this app sharing your location data for the past two weeks \textbf{publicly}?

Q23: Consider all the apps you regularly use on your phone (not just the apps listed earlier). Which app would you \textbf{most} trust to use your location data for the purposes of studying and mitigating COVID-19? Why? [free response]

Q24: Which app that you currently use would you \textbf{least} trust to use your location data for the purposes of studying and mitigating COVID-19? Why? [free response]

\subsection{Current use of COVID-19 app}

Q25: Have you used any apps that help track the spread of COVID-19? (for example, Singapore's ``TraceTogether'') [yes, no]

\underline{If yes, participants branch to `already have app.'}

\underline{If no, participants continue.}

\subsection{New app, perfect privacy}

\textit{Imagine there is a \textbf{new} app that would track your location at all times for the purposes of mitigating the spread of COVID-19.}
 
\textit{Suppose that this app protects your data perfectly.}
 
Q50: How likely would you be to install and use this app?

Q51: Would this app change your current behavior?

Q52: Optionally, please use this space tell us any initial thoughts you have about such an app. [free response]

\subsection{New app, app makers know location}
\textit{Imagine there is a new app that would track your location at all times for the purposes of studying or mitigating the spread of COVID-19.} \label{protocol:new-app}

\textit{Suppose now that the makers of the application would know your location at all times, but would not share your location with any other entity. }
 
Q55: How likely would you be to install and use this app?

Q56: Now, suppose that the app is made by one of the following companies, all of which already have created popular apps. Please rate how comfortable you would be if each company were responsible for this new app. [``I don't know enough about this company to make a  decision'' + 5-pt Likert scale for each of the following]
\begin{itemize}
    \item Google (Google Maps, Waze, etc)
    \item Apple (Apple Maps)
    \item Facebook (Facebook, Facebook Messenger, Instagram, WhatsApp)
    \item Microsoft (Skype, OneDrive, etc)\footnote{Added April 17 (week 3).}
    \item ByteDance (TikTok)
    \item Zoom Video Communication
    \item Uber
    \item Lyft
    \item AirBnb
    \item MyFitnessPal
    \item AllTrails
    \item FitBit
\end{itemize}

Q57: Suppose that the app is made by one of the following general entities. Please rate how comfortable you would be if one of the following were responsible for this new app, which would use the location data they collect from your smartphone to track the spread of COVID-19. [5-pt Likert scale for each of the following]
\begin{itemize}
    \item A university research group
    \item An activist group 
    \item An industry startup
    \item Your government
    \item The United nations
\end{itemize}

Q58: Optionally, please use the space below to elaborate on your thoughts about one or more companies using your location data for the purposes of tracking COVID-19. [free response]

\subsection{New app, app makers share data with government}

\textit{Again, imagine there is a new app that would track your location at all times for the purposes of studying or mitigating the spread of COVID-19.}
 
\textit{Suppose now that the makers of the application would know your location at all times, and would also share that data with your government if you were diagnosed with COVID-19.}

Q61: How likely would you be to install and use this app?

Q62: If the government’s use of the data were \textbf{supervised by a judge}, how likely would you be to install and use this app?\footnote{This question, and the rest of this section, was added on April 17 (week 3) as a previous version was ambiguous.}

\textit{Now suppose that the makers of the application would share your location data with your government \textbf{only if you tested positive for COVID-19.}}

Q63: How likely would you be to install and use this app?

Q112: If the government’s use of the data were \textbf{supervised by a judge}, how  likely would you be to install and use this app?

Q64: Optionally, do you have any other thoughts about a company that is doing COVID-19 tracking sharing your location with your government? [free response]

Q115:  Optionally, do you have any other thoughts about judicial oversight of the government’s usage of location data? [free response]

\subsection{Other location data sources: Credit card history and surveillance camera footage}\label{protocol:other-loc}

\textit{There\footnote{Added April 17 (week 3)} are other ways to track someone's location. One is the use of video cameras in public places. Another is the use of credit card purchasing histories.}

Q115: How comfortable would you be with your \textbf{credit card company} deriving your location history for the past two weeks for the purposes of studying and mitigating the spread of COVID-19?

Q116: How comfortable would you be with your \textbf{credit card company} deriving your location history for the past two weeks \textbf{and sharing it with your government} for the purposes of studying and mitigating the spread of COVID-19?

Q118: Optionally, do you have any other thoughts about your location history being derived from your \textbf{credit card} purchase history? [free response]

Q117: How comfortable would you be with \textbf{your government} deriving your location history for the past two weeks from \textbf{surveillance camera footage} for the purposes of studying and mitigating the spread of COVID-19? 

Q119: Optionally, do you have any other thoughts about your location history being derived from \textbf{surveillance camera footage}? [free response]

\subsection{Other data location sources II: wearable electronics and public area sensors}

Suppose\footnote{Added July 17 (week 16)} there were an electronic bracelet that would track your location for the purposes of studying or mitigating the spread of COVID-19.

Q146: How likely would you be to use this bracelet?

Q147: Optionally, do you have any other thoughts about wearable electronics being used for the purposes of studying or mitigating the spread of COVID-19? [free response]

Suppose your region added sensors (such as cameras, phone tagging stations, etc) in public areas (such as subway stations, bus stops, storefronts, public parks, etc). 

Q148: How comfortable would you be with the use of public-area sensors to study or mitigate the spread of COVID-19?

Q149: Optionally, do you have any other thoughts about such sensors being used for the purposes of studying or mitigating COVID-19? [free response]

\subsection{Proximity tracing}
\label{protocol:prox}
\textit{One alternative\footnote{Added April 17 (week 3)} to location tracking for the purposes of studying or mitigating COVID-19 is \textbf{proximity tracing}, in which your phone would automatically exchange information with every phone within 6 feet (2 meters) of your phone, \textbf{keeping track of your close physical encounters, but not tracking your actual location}. This data could then be used to reconstruct your close encounters if you contracted COVID-19, or could alert you if someone you had been in close physical proximity to tested positive for COVID-19.}

Q121: Imagine that your \textbf{cell phone manufacturer or phone operating system} would conduct proximity tracing for the purposes of studying or mitigating COVID-19 (and, vice versa, other phones will record that they have been in the proximity of your phone). How comfortable would you be with this?

Q122: Suppose that your \textbf{cell phone manufacturer or phone operating system} would share this proximity data \textbf{with your government} if you tested positive for COVID-19. How comfortable would you be with this?

Q123: Optionally, do you have any other thoughts about your cell phone manufacturer or phone operating system tracking other phones nearby? [free response]

\textit{Imagine instead there is a new \textbf{app} that would conduct proximity tracing for the purposes of studying or mitigating COVID-19: that is, it would not track your location, but would instead keep track of other phones that you are nearby (and, vice versa, other phones with this app will record that they have been in the proximity of your phone).}

Q124: How likely would you be to download this app?

Q125: Now, suppose that the proximity tracing app is made by one of the following companies. Please rate how comfortable you would be if each company were responsible for this new app.
[``I don't know enough about this company to make a  decision'' + 5-pt Likert scale for each of the following]
\begin{itemize}
    \item Google (Google Maps, Waze, etc)
    \item Apple (Apple Maps)
    \item Facebook (Facebook, Facebook Messenger, Instagram, WhatsApp)
    \item Microsoft (Skype, etc)
    \item ByteDance (TikTok)
    \item Zoom Video Communication
    \item Uber
    \item Lyft
    \item AirBnb
    \item MyFitnessPal
    \item AllTrails
    \item FitBit
\end{itemize}

Q126: Now, suppose that the proximity tracing app is made by one of the following general entities. Please rate how comfortable you would be if each entity were responsible for this new app.
\begin{itemize}
    \item A university research group
    \item An activist group 
    \item An industry startup
    \item Your government
    \item The United nations
\end{itemize}

Q127: Optionally, do you have any other thoughts about an app that tracks other phones nearby?

Q128: Optionally, do you have any other thoughts about proximity tracking versus location tracking for the purposes of studying or mitigating COVID-19?

\subsection{Government use of data}
\textit{\textbf{If the government acquired your location data or proximity data\footnote{'or proximity data' added April 17}}(i.e. from an app on your phone, from your cell phone carrier, etc) \textbf{for the purposes for studying and mitigating COVID-19}....}
 
Q66: How likely do you think it is that your government would \textbf{delete} the data after the pandemic ends?

Q67: How likely do you think it is that your government would \textbf{only} use the data for the purposes of tracking COVID-19?

Q68: How concerned would you be about your government’s use of your location data harming \textbf{your personal safety} or the safety of those in your community? 

Q69: Suppose your location was shared with only a specific sector of the government. For each of the following sectors of government, please rate how comfortable you would be with them having access to your location data. 
\begin{itemize}
    \item Federal Disease Tracking Agency (US: CDC)
    \item Your state or City Health Department
    \item Tax Agency (US: IRS)
    \item Local law enforcement (state, country, city, etc)
    \item Immigration authorities (US: CBP or ICE)
\end{itemize}

Q70: Optionally, please use the space below to elaborate on your thoughts about the government having access to your location data for the purposes of COVID-19 tracking.  [free response]

\subsection{App features}
\textit{In some countries, such as South Korea, China, and Singapore, there do exist apps to monitor the spread of COVID-19 through location tracking. These apps can have multiple purposes, including:}

- \textit{Alerting the user if they have come into contact with someone who later tests positive with COVID-19;}

- \textit{Helping the community or law enforcement enforce isolation and quarantine edicts;}

- \textit{Tracing viral strains through the community.}
 
Q72: If a new app were deployed in your country to mitigate the spread of COVID-19, which of the following features would you want it to have? (5-point Likert-scale for each of the following:)
\begin{itemize}
    \item Notify you if you came close to someone who later tested positive for COVID-19
    \item Notify anyone you came close to in the past two weeks if you tested positive for COVID-19
    \item Make your location history for the past two weeks publicly available if you tested positive for COVID-19
    \item Make public a database of the location histories of anyone who tested positive for COVID-19
    \item Notify you if your neighbors were not isolating themselves as recommended or mandated
    \item Let you notify the authorities if you saw people you suspected or knew to be breaking the isolation recommended or mandated
    \item Automatically notify the authorities if people were not isolating as mandated
    \item Used by scientists to study tends, not individuals
    \item Geofence to enforce mandatory or voluntary quarantine
    \item General assessment of social distancing in an area to display areas of high congregation
\end{itemize}

Q73: Optionally, do you have any other thoughts about what you would want such an app to do? [free response]

Q74: Optionally, do you have any other thoughts about what you would want such an app to NOT do? [free response]
 
Q75: Is such an app available in your country? [Yes / No / I’m not sure]
\underline{If yes, participants branch to ‘App available’}

\subsection{Section 5.0: Prior privacy preferences}

\textit{We’re now going to ask you about your thoughts about location sharing with your government BEFORE COVID-19.}
 
Q80: In Oct 2019 (before the first known cases of COVID-19), how comfortable would you have been with your location data being shared with the government in general?

Q81: In Oct 2019 (before the first known cases of COVID-19), how comfortable were you with your location data being shared with the following sectors of government? [5-Point Likert scale for each of the following:]
\begin{itemize}
    \item Federal Disease Tracking Agency (US: CDC)
    \item Your state or City Health Department
    \item Tax Agency (US: IRS)
    \item Local law enforcement (state, country, city, etc)
    \item Immigration authorities (US: CBP or ICE)
\end{itemize}

Q82: Optionally, please use the space below to elaborate on your thoughts about one or more companies sharing your location data with some part of the government (before COVID-19). [free response]

\subsection{Demographics II}

Almost done!

Q39: What is your age? (you may answer approximately if you do not know, or wish not to say exactly) [free response]

Q40: What is your gender identity? [free response]

Q141: Please select any races or ethnicities that you feel accurately reflect who you are. Please select as many as apply to you. We also realize that because race and ethnicity cannot be put into categories, you may prefer to self-describe your race and ethnicity in the following question. You may also select from the options below and submit a free-response. The following races and ethnicities are presented in alphabetical order. [American Indian or Alaskan Native, Asian, Black or African American, Hispanic, Native Hawaiian or Other Pacific Islander, White]\footnote{Added week 12}

Q142: If you prefer to self-describe your race and ethnicity instead of or in addition to using the checkboxes above, please do so here. [free response]\footnote{Added week 12}

Q41: What political party do your views typically align with? [free response]

Q42: What are your top three news sources? (i.e. Twitter, Facebook, Fox News, CNN, NPR, New York Times, etc) [free response]

Q132: Have you ever had COVID-19? [Yes, definitely; Yes, I think so; I am unsure; No, I don't think so; No, definitely not]\footnote{Added week 5}

Q133: Have you ever been medically tested for COVID-19? [Yes, I have had a test; No, I have not been tested]\footnote{Added week 5}

Q43: Regarding COVID-19, are you in high risk group or live with someone with high risk? [yes / no]

Q44: Are you generally interested in or concerned about privacy and technology? [yes / no]

Q45: Do you know how to change location permissions for apps on your phone? [yes / no]

Q129: What is your phone manufacturer? (e.g. Apple, Samsung, Huawei, Nokia, OnePlus, XiaoMi, OPPO, etc) [free response]\footnote{Added week 5} 

\subsection{Branch: app is available}

Q76: Have you downloaded the app in your country to mitigate or study the spread of COVID-19? [yes, no]

Q77: If you have not downloaded the app: why not? what changes, or assurances by the manufacturer or government (if any) would you want to see to the app before downloading? [free response]

Q78: What are your thoughts about the privacy properties of this app? [free response]

\subsection{Branch: Already have app}

You indicated that there is an app (or apps) available in your country to track, study, or mitigate COVID-19. This section will ask about that app.

Q27: What is the name of the app, or apps?

Q28: Why did you install and use it?

Q29: Do you know anyone who did not download the app? [yes, no]

Q30: If so, why did they not install it?

Q31: What concerns, if any, do you have about the app?

Q32: If you had or have concerns, what outweighed the concerns and lead you to the decision to download the app?

Q33: What concerns, if any, do you have about your government having access to your location data?

Q34: What concerns, if any, do you have about the app makers having access to your location data?

Q35: Do you expect the app makers to stop storing your location data after the pandemic is over?

Q36: Do you plan to delete the app after the pandemic?

Q37: Anything else you’d like to say about the app and/or your concerns?

\end{document}